\documentclass[preprint]{aastex}
\usepackage{mkfig}

\begin{document}
 
\title{\bf Two New Low Galactic D/H Measurements from FUSE\altaffilmark{1}}

\author{Brian E. Wood\altaffilmark{2}, Jeffrey L. Linsky\altaffilmark{2},
  Guillaume H\'{e}brard\altaffilmark{3,4},
  Gerard M. Williger\altaffilmark{4}, H. Warren Moos\altaffilmark{4},
  William P. Blair\altaffilmark{4}}

\altaffiltext{1}{Based on observations made with the NASA-CNES-CSA Far
  Ultraviolet Spectroscopic Explorer.  FUSE is operated for NASA by the
  Johns Hopkins University under NASA contract NAS5-32985.}
\altaffiltext{2}{JILA, University of Colorado and NIST, Boulder, CO
  80309-0440; woodb@origins.colorado.edu, jlinsky@jila.colorado.edu.}
\altaffiltext{3}{Institut d'Astrophysique de Paris, CNRS, 98 bis Boulevard
  Arago, F-75014 Paris, France; hebrard@iap.fr.}
\altaffiltext{4}{Department of Physics and Astronomy, Johns Hopkins
  University, 3400 North Charles Street, Baltimore, MD 21218;
  williger@pha.jhu.edu, hwm@pha.jhu.edu, wpb@pha.jhu.edu.}

\begin{abstract}

     We analyze interstellar absorption observed towards two subdwarf O
stars, JL~9 and LSS~1274, using spectra taken by the {\em Far Ultraviolet
Spectroscopic Explorer} (FUSE).  Column densities are measured for many
atomic and molecular species (H~I, D~I, C~I, N~I, O~I, P~II, Ar~I, Fe~II,
and H$_{2}$), but our main focus is on measuring the D/H ratios
for these extended lines of sight, as D/H is an important diagnostic for
both cosmology and Galactic chemical evolution.  We find
${\rm D/H}=(1.00\pm 0.37)\times 10^{-5}$ towards JL~9, and
${\rm D/H}=(0.76\pm 0.36)\times 10^{-5}$ towards LSS~1274 (2$\sigma$
uncertainties).  With distances of $590\pm 160$~pc and $580\pm 100$~pc,
respectively, these two lines of sight are currently among the longest
Galactic lines of sight with measured D/H.  With the addition of these
measurements, we see a significant tendency for longer Galactic lines of
sight to yield low D/H values, consistent with previous inferences about the
deuterium abundance from D/O and D/N measurements.  Short lines of sight
with H~I column densities of $\log N({\rm H~I})<19.2$ suggest that the
gas-phase D/H value within the Local Bubble is
${\rm (D/H)_{LBg}}=(1.56 \pm 0.04) \times 10^{-5}$.
However, the four longest Galactic lines of sight with measured D/H, which
have $d>500$~pc and $\log N({\rm H~I})>20.5$, suggest a significantly
lower value for the true local-disk gas-phase D/H value,
${\rm (D/H)_{LDg}}=(0.85\pm 0.09)\times 10^{-5}$.  One interpretation of
these results is that D is preferentially depleted onto dust grains relative
to H and that longer lines of sight that extend beyond the Local Bubble
sample more depleted material.  In this scenario, the higher Local Bubble
D/H ratio is actually a better estimate than ${\rm (D/H)_{LDg}}$ for the true
local-disk D/H, ${\rm (D/H)_{LD}}$.  However, if
${\rm (D/H)_{LDg}}$ is different from ${\rm (D/H)_{LBg}}$ simply because of
variable astration and incomplete ISM mixing, then
${\rm (D/H)_{LD}}={\rm (D/H)_{LDg}}$.

\end{abstract}


\section{INTRODUCTION}

     One of the great successes of the Big Bang theory is that it predicts
the light element abundances in the universe with reasonable accuracy.
However, the exact abundance predictions depend on the cosmic baryon
density, $\Omega_{b}$.  The abundance most sensitive to $\Omega_{b}$ is
deuterium, so measuring the primordial deuterium-to-hydrogen ratio has
become an important method for constraining $\Omega_{b}$ in cosmological
models \citep[e.g.,][]{amb85,sb01}.
A good estimate of the primordial D/H ratio can be obtained from IGM
absorption components seen towards distant quasars, since the metal
abundance in the IGM is small, indicating that the gas has experienced very
little astration.  \citet{dk03} compile
several such measurements \citep[e.g.,][]{jmo01,mp01,sal02}
and report a primordial D/H of
${\rm (D/H)_{prim}}=(2.78^{+0.44}_{-0.38})\times 10^{-5}$, although there
is significant scatter in the individual measurements\footnote{Unless
otherwise noted the quoted uncertainties for averaged quantities are
1$\sigma$ standard deviations of the mean, but quoted uncertainties for
individual measurements are 2$\sigma$.}.  This value is
consistent with primordial deuterium abundances inferred from recent
WMAP measurements of the cosmic microwave background, combined with
Big Bang nucleosynthesis calculations \citep{dr03}.

     Since deuterium is gradually destroyed in the interiors of stars, it
is expected that the D/H value should be lower in places that have
experienced a lot of stellar processing.  \citet{krs04}
find a value of ${\rm D/H}=(2.2\pm 0.7)\times 10^{-5}$ for Complex~C,
a high velocity cloud falling onto the Milky Way, which has a low
metallicity but presumably has experienced more stellar processing than
most IGM material.  The Complex~C D/H measurement may be slightly lower
than the ${\rm (D/H)_{prim}}$ value quoted above,
although the uncertainties are too large to provide complete confidence
in this result.  The D/H ratio in the local disk region of
our galaxy, ${\rm (D/H)_{LD}}$, should be significantly lower than both the
Complex~C and IGM measurements due to significantly more stellar processing.

     Comparing ${\rm (D/H)_{LD}}$ to ${\rm (D/H)_{prim}}$ provides an
excellent indication of the amount of stellar processing experienced by
interstellar material in our Galaxy, and therefore provides a useful test
of Galactic chemical evolution models.  Many measurements have been made of
the gas-phase local-disk D/H ratio in the Galaxy [${\rm (D/H)_{LDg}}$].
These measurements do not account for deuterium that may be locked into
molecules and dust, but it has nevertheless been assumed in the past that
${\rm (D/H)_{LDg}}\approx {\rm (D/H)_{LD}}$.  Unfortunately, the D/H
measurements for various Galactic lines of sight have not collectively
provided an unambiguous value for ${\rm (D/H)_{LDg}}$.  There seems to be
substantial variation in the ${\rm (D/H)_{LDg}}$ measurements, at least for
longer lines of sight.  However, the D/H measurements at least all
suggest that ${\rm (D/H)_{LDg}}<{\rm (D/H)_{prim}}$ as one would expect
\citep[e.g.,][]{jll98,hwm02}.

     Ultraviolet spectra of interstellar H~I and D~I Lyman-$\alpha$
absorption lines from the {\em Hubble Space Telescope} (HST) have provided
many accurate Galactic D/H measurements
\citep[e.g.,][]{jll95,bew96,ard97,np97}.  However, the D~I Ly$\alpha$
line merges with the H~I line for H~I column
densities above $\sim 5\times 10^{18}$ cm$^{-2}$.  Thus, the HST
results only apply for short lines of sight with low H~I column densities.
In particular, the HST measurements do not reach beyond the boundaries of
the Local Bubble, which is the region within $\sim 100$~pc of the Sun
in which most of the volume consists of very hot, low density, ionized ISM
material \citep[e.g.,][]{sls98,dms99,rl03}.

     Analogous to the ${\rm (D/H)_{LD}}$ and ${\rm (D/H)_{LDg}}$ quantities
defined above, we define ${\rm (D/H)_{LB}}$ and ${\rm (D/H)_{LBg}}$ to be
the total and gas-phase D/H values for the Local Bubble, respectively.
\citet{jll98} quotes a value of
${\rm (D/H)_{LBg}}=(1.5\pm 0.1)\times 10^{-5}$ based on HST measurements of
the Local Interstellar Cloud (LIC) immediately surrounding the Sun.  This
is about a factor of 2 lower than the ${\rm (D/H)_{prim}}$ value quoted
above.  More recent measurements from the {\em Far Ultraviolet
Spectroscopic Explorer} (FUSE) have confirmed this result
\citep{hwm02,cmo03}.  There is no convincing evidence for D/H
variations within the Local Bubble, but ${\rm (D/H)_{LBg}}$ will equal
${\rm (D/H)_{LDg}}$ only if interstellar gas in the Galaxy is relatively
homogeneous and well mixed.  Consideration of longer lines of sight is
necessary to determine if ${\rm (D/H)_{LDg}}={\rm (D/H)_{LBg}}$.

     Measuring D/H for longer lines of sight requires access to Lyman
lines higher than Ly$\alpha$.  Observations of these lines have been
provided by the {\em Copernicus} satellite in the 1970s, the Interstellar
Medium Absorption Profile Spectrograph (IMAPS) instrument (part of the
{\em ORFEUS-SPAS~II} experiment onboard the STS-80 Space Shuttle Columbia
flight in 1996), and more recently by FUSE, which was launched in 1999.
Measurements of D/H from these instruments
\citep[e.g.,][]{dgy76,avm77,gs00,cgh03}
suggest a significant amount of variability for D/H outside the Local
Bubble within the range ${\rm D/H}=(0.5-2.2)\times 10^{-5}$ \citep{hwm02}.
When combined with the ${\rm (D/H)_{prim}}$ measurement, this
suggests a deuterium destruction factor of $\sim 1.3-5.6$ in the local disk
region of the Galaxy.  The actual average ${\rm (D/H)_{LDg}}$ value and
the average deuterium destruction factor are presumably somewhere in the
middle of the above ranges, but it is not yet clear exactly where.

     The spatial variability of the Galactic D/H should disappear if one
considers only lines of sight longer than the scale size of the variations,
so the longest and highest column density lines of sight should in
principle provide the best measurements of ${\rm (D/H)_{LDg}}$.  In
practice, however, the high column densities for long Galactic lines of
sight often lead to a very confusing spectrum of blended atomic and
molecular absorption lines, making identification and accurate measurement
of deuterium lines impossible in many cases.  Nevertheless, considering
D/O and D/N measurements with previously published D/H values,
\citet{gh03} suggest that the deuterium abundance for long
lines of sight is significantly lower than in the Local Bubble.  However,
there are very few measurements for distances longer than 500~pc.  In this
paper, we will double the number of D/H measurements in this distance
regime by analyzing FUSE observations of two new lines of sight with
$d>500$~pc.

\section{THE TARGETS}

     Our target stars for this project are JL~9, which acquires its name
from the catalog of \citet{sj74}, and LSS~1274, which acquires
its identifier from a catalog of luminous southern hemisphere stars by
\citet{cbs71}.  The properties of these two stars are
listed in Table~1.  Both are subdwarf~O (sdO) stars.  Hot subdwarf stars
are excellent targets for our purposes because they provide strong UV
continua against which many ISM absorption lines can be seen (including
H~I and D~I lines).  They are significantly brighter than white dwarfs and
can therefore be used to observe longer lines of sight, but they are not
as bright as OB main sequence stars, which are often too bright to observe
with FUSE's sensitive detectors, even at large distances.

     \citet{sd93} computes a distance of $d=580\pm 100$~pc for
LSS~1274 from spectroscopic analysis, but unfortunately there is no
similar measurement for JL~9.  A distance for JL~9 can be estimated
assuming it has an absolute magnitude similar to other sdO stars with
measured distances.  In particular, we use the hot subdwarf catalog of
\citet{dk88} to identify all sdO stars with $V<11$
that may be bright and near enough to have reasonably accurate
{\em Hipparcos} distance measurements.  We then use the SIMBAD database to
determine which stars indeed have {\em Hipparcos} distances.  Only seven
stars meet these criteria (CD-31~1701, BD+75~325, Feige~66, HD~127493,
HD~149382, BD+28~4211, and BD+25~4655).  After adding LSS~1274 and its
spectroscopic distance to this group, we find an average absolute
magnitude of $M_{V}=4.40\pm 0.58$ for this selection of sdO stars.
This average sdO magnitude is very similar to the value that \citet{pt97}
find for the more numerous sdB stars.  Assuming that this is a reasonable
estimate for JL~9 yields a distance of $d=590\pm 160$~pc, which is the
distance reported in Table~1.

\section{THE FUSE OBSERVATIONS}

     Table~2 lists the FUSE observations used in our analysis.  JL~9 was
observed once through the low-resolution (LWRS) aperture, while LSS~1274
was observed three separate times through the medium-resolution (MDRS)
aperture.  The JL~9 observation was made in a single 16.6~ksec
exposure, while the LSS~1274 data were taken in 77 separate exposures
totaling 85.0~ksec.

     \citet{gh03} have already processed the LSS~1274
observations and analyzed them to some extent (see \S4.1).  We use the
same processed data set.  The JL~9 data are processed similarly, using
CALFUSE v2.4.1.  The FUSE instrument uses a multi-channel design to fully
cover its 905--1187~\AA\ spectral range --- two channels (LiF1 and LiF2)
use Al+LiF coatings, two channels (SiC1 and SiC2) use SiC coatings, and
there are two different detectors (A and B).  For a full description of the
instrument, see \citet{hwm00} and \citet{djs00}.  With this
design FUSE acquires spectra in eight segments covering different,
overlapping wavelength ranges.  For LSS~1274, the numerous
exposures are cross-correlated and coadded to produce a single spectrum
for each segment.  For JL~9, this is not absolutely necessary since the
observation was taken in a single exposure.  However, by first breaking
the time-tagged photon address mode observation into 17 subexposures, and
then cross-correlating and coadding the resulting 17 spectra, we found we
could noticeably improve the spectral resolution of the JL~9 data,
particularly for the SiC segments.  Thus, these are the spectra that are
principally used in our analysis.

     Although we coadd the individual exposures of the FUSE data set, we
do not coadd the eight overlapping spectral segments that result from the
data reduction, due to concerns that such an operation could significantly
degrade the resolution.  In Figure~1, we display the full FUSE spectrum
of JL~9, which is constructed by splicing together the following segments:
SiC1B ($912-918$~\AA), SiC2A ($918-996$~\AA), LiF1A ($996-1080$~\AA),
SiC2B ($1080-1090$~\AA), and LiF1A ($1090-1180$~\AA).  The spectrum is
riddled with numerous atomic and H$_{2}$ absorption lines from the ISM.
The LSS~1274 spectrum is quite similar to that of JL~9 shown in Figure~1,
and has a similar selection of absorption lines.  This is not surprising
given that both are sdO stars at roughly the same distance (see Table~1).

     There is no wavelength calibration lamp on FUSE, so determining the
absolute wavelength scale can be difficult.  We use the numerous
H$_{2}$ lines seen throughout the spectrum as wavelength calibration lines
to ensure that the full spectrum is on a self-consistent relative
wavelength scale.  In other words, we calibrate the wavelengths to ensure
that all H$_{2}$ lines are centered on the same velocity.  For JL~9, the
implied wavelength corrections are relatively uniform across the individual
spectral segments, so a single wavelength correction factor is used for
each entire segment.  However, for LSS~1274 the wavelength corrections
implied by the H$_{2}$ lines vary significantly (up to $\sim 5$ km~s$^{-1}$)
within several of the segments, forcing us to use variable correction
factors across the segments.  For the {\em absolute} wavelength calibration,
we force the H$_{2}$ line velocities to agree with the average H$_{2}$ line
velocity seen within the LiF1A segment.  We choose LiF1A to establish the
absolute wavelength scale because the LiF1 channel is used for target
guiding, meaning that the target should be most accurately centered in the
aperture for the LiF1A and LiF1B segments, and those segments should
therefore have the best absolute wavelength scales.

     Geocoronal airglow emission is observed within the strongest of the
broad H~I lines (e.g., H~I Ly$\beta$, Ly$\gamma$, and
Ly$\delta$ at 1025.7~\AA, 972.5~\AA, and 949.7~\AA, respectively).
This emission is naturally stronger in the JL~9 spectrum, since those data
were taken through the LWRS aperture rather than the narrower MDRS aperture.
Fortunately, for both stars the airglow lines are roughly centered
in the saturated cores of the broad H~I absorption lines, so the emission
can be subtracted with reasonable accuracy.  Figure~1 shows the JL~9
data after this subtraction has already been performed.  Nevertheless,
for JL~9 there may be flux inaccuracies near the short wavelength side of
the H~I Ly$\gamma$ and Ly$\delta$ lines due to uncertainties in the
airglow removal.  No attempt is made to correct for O~I and N~I airglow
emission.  However, this emission is potentially a problem for only a
small number of lines, and probably not a problem at all for the MDRS
LSS~1274 data.  Nevertheless, the excess emission in the bottom of the
JL~9 O~I absorption feature at 988.4~\AA\ in Figure~1 is surely airglow
emission.

\section{ANALYSIS}

\subsection{Profile Fitting Procedures}

     Figure~1 shows that the FUSE spectra contain numerous interstellar
lines of 13 different atomic species, and a forest of H$_{2}$ lines from
rotational levels $J=0-5$.  The primary goal in analyzing these data is
to extract accurate column densities from these absorption lines for all
of the atomic and molecular species represented in the spectra.  The
final results are listed in Table~3 and we describe below the profile
fitting methods used to measure these column densities and 2$\sigma$
uncertainties.  Note that Table~3 does not list column densities for C~II,
C~III, N~II, N~III, and Si~II.  There are lines of these species in the
spectra and they are even included in the fit in Figure~1, but we do not
believe that we can derive reliable column densities from these lines.
There are several reasons for this.  One is that each of these species is
represented by only one or two highly saturated lines in the flat part of
the curve of growth that cannot yield precise column density
measurements.  The C~II, N~III, and Si~II lines are also highly blended
with other lines.  Finally, we are concerned about the possibility of
stellar absorption contaminating the interstellar C~II, C~III, N~II, and
N~III lines, especially for LSS~1274.

     Our primary approach to measuring column densities is to perform
global fits to the absorption lines, like the fit shown in Figure~1.  In
this approach, a spectrum covering the full FUSE spectral range is
constructed from the FUSE segments, as described in \S3, and all the lines
in the spectrum are then fitted simultaneously.  This approach has been
previously used to analyze the lines of sight to WD~1634-573 and
WD~2211-495 \citep{bew02,gh02}.  However, as
in these two previous cases we also perform an independent analysis using
the Owens.f profile fitting code, which was developed by the French FUSE
team and has been used extensively for measuring ISM lines in FUSE
observations of other targets
\citep{sdf02,jwk02,nl02,ml02,gs02,gh03,cgh03,cmo03}.

     Column density measurements from absorption lines are susceptible to
many systematic errors:  continuum placement issues, unresolved velocity
structure, unidentified blends, uncertain instrumental line profile, etc.
Performing two independent and very different analyses allows us to see
whether different assumptions and methods still lead to similar
measurements.  This ultimately increases confidence in our final results.
The Owens.f analysis provides column density measurements for the
important D~I, O~I, N~I, and Fe~II species, so the column densities
reported in Table~3 for these species are compromises between the results
of the two independent analyses.  The Owens.f analysis for LSS~1274 has
already been reported by \citet{gh03}, and we refer the
reader to that paper for details.  A very similar analysis is performed on
the JL~9 data, but we focus here mostly on the global fitting method.

     The first step in performing a global fit to a spectrum like that
in Figure~1 is to estimate the stellar continuum.  This is initially done
with the help of a polynomial fitting routine to extrapolate over the
absorption lines, although the continuum is refined after initial fits
to the data in order to improve the quality of the fit.  We do not try to
compute synthetic model continua for these stars, because their stellar
parameters are poorly known (especially JL~9), and because such models are
of limited accuracy in matching observations \citep[e.g.,][]{sdf02}.
The entire spectrum and all the ISM absorption lines within it are fitted
simultaneously.  We use the \citet{dcm03} list of atomic lines as our
source for all necessary atomic data.  We include in our fit all
reasonably strong lines of the 13 atomic species with at least one
detected line (see Fig.~1).  This includes not only lines that are clearly
detected but many undetected lines as well, since nondetections can also
provide constraints for the fit.  We use the \citet{ha93a,ha93b}
lists of Lyman and Werner band H$_{2}$ lines to fit the H$_{2}$ absorption.
We include all $v=0$, $J=0-5$ transitions in the fit, once
again including both detected and undetected lines.  The final tally of
lines included in the fit is 140 atomic and 312 H$_{2}$ lines.  Their
locations are shown in Figure~1.

     The three parameters of any absorption line fit are the line centroid
velocity ($v$), the column density ($N$), and the Doppler broadening
parameter ($b$).  In our fits, all lines of a given atomic species are
naturally forced to have the same column density.  The atomic lines are
also forced to have the same centroid velocities, except for H~I.  The H~I
lines are much stronger than the other lines and are therefore sensitive
to weaker ISM velocity components.  They clearly have a different centroid
velocity because of this, so in the fits the H~I lines are allowed to
have their own independent velocity.  The Doppler parameter (in km~s$^{-1}$)
is related to the temperature ($T$) and nonthermal velocity ($\xi$) of the
ISM gas by $b^{2}=0.0165T/A+\xi^{2}$, where $A$ is the atomic weight of
the species in question.  In our fits, we use $T$ and $\xi$ as
free parameters and compute $b$ values for all atomic lines from these
parameters.  In this way, the fits to all individual atomic lines and the
fit parameters are in some sense interdependent.  However, the H~I lines are
once again treated separately and
allowed to have their own independent $b$ value.  The H$_{2}$ lines are
surely formed in different, cooler regions of the ISM than the atomic
lines, so their velocities and Doppler parameters are allowed to be
different from those of the atomic lines.  All H$_{2}$ lines of a given
rotational level are naturally forced to have the same column density.

     Spectral regions that are heavily contaminated by stellar absorption
are ignored in the fits.  Examples in Figure~1 include the
$922.0-924.5$~\AA, $933.3-933.9$~\AA, $944.5-945.0$~\AA, $955.3-955.6$~\AA,
$1031.5-1032.5$~\AA, and $1037.5-1038.0$~\AA\ regions, which are
contaminated by stellar absorption lines of N~IV, S~VI, and O~VI.
We also ignore the region around the O~I $\lambda$988 line since it is
contaminated by airglow emission (see \S3).

     The spectral resolution of FUSE is not sufficient to resolve narrow
ISM absorption lines, so a fit to the data must be convolved with the
instrumental line spread function (LSF) before being compared to the
data.  Unfortunately, the FUSE LSF is not well known \citep{jwk02}.
It varies with wavelength, it can vary from one observation to the next,
and it can also depend on exactly how the data are processed (see \S3).
For these reasons, the LSF is generally a free parameter of our fits, where
we assume a 2-Gaussian representation for the LSF.  However, we make no
attempt to correct for variations of the LSF with wavelength, which is a
potential source of systematic error in the analysis.  \citet{bew02}
derived an average FUSE LSF from various fits to FUSE spectra.  We
experiment with simply using that LSF in our analysis, although we find
that the LSFs of our spectra are somewhat narrower than this,
especially for the LSS~1274 data.

     Figure~1 shows one fit to the JL~9 data, but ultimately a large number
of fits are considered before estimating best values and uncertainties for
the ISM column densities.  For example, we experiment with different
continuum estimations.  This is done in several different ways, but the
simplest is to arbitrarily increase or decrease the entire continuum by
various percentages to see if the fits to the various absorption lines
still look reasonable and to see how much the derived column densities
change as a consequence of the continuum variation.  As mentioned above,
we also experiment with using the LSF from \citet{bew02} rather than
allowing the LSF to vary.

     Although we generally work with FUSE spectra constructed from the
individual FUSE segments as described in \S3, we also try using the SiC1B
segment to cover the entire region below 990~\AA\ instead of using SiC2A.
The extra attention to this spectral region is warranted since all of the
important D~I, N~I, and O~I lines are located below 990~\AA.

     Most of the observed N~I and O~I lines are saturated and lie in the
flat part of the curve of growth.  In order to make sure that including
these lines in the fit is not leading to column densities radically
different from those suggested by the optically thin lines, we experiment
with fits in which the saturated N~I and O~I lines are ignored.

     Finally, we also experiment with two-component fits to the data
rather than single-component fits.  Unfortunately, FUSE does not have
sufficient resolution to assess the velocity structure of the ISM along
our two lines of sight, and no other high resolution observations of JL~9
or LSS~1274 exist that can assist us in this matter.  This lack of
knowledge of the ISM structure is a potentially significant source of
systematic error, so we try fits to the data with two components to test
whether these fits result in significantly different column densities.  For
example, we try fits with components that have a velocity separation of
10 km~s$^{-1}$ and a column density difference in all lines of 0.5 dex.

     The idea behind all this experimentation is to collect a large number
of acceptable fits to the data with different assumptions, and then for
each column density in question we use the range of values suggested by
this set of fits to define the best value and its uncertainty.
Uncertainties derived in this fashion are not statistical in nature, but
we believe they can be considered approximately 2$\sigma$ errors, in the
sense of representing roughly 95\% confidence intervals.  The uncertainties
are certainly larger than 1$\sigma$ since we do not throw out 32\% of our
acceptable fits in estimating the errors.

     As mentioned above, the results of the global fits are compared with
those of the Owens.f code (at least for D~I, N~I, O~I, and Fe~II), and
the final column densities and uncertainties reported in Table~3 are
essentially averages of the two independent measurements.  The two
analyses yield column densities that generally agree very well.  The
only exception is the N~I column density for LSS~1274, which is
discussed in \S4.3.4.  The Owens.f analysis represents a significantly
different approach to fitting the data in terms of line selection,
continuum estimation, LSF treatment, and uncertainty derivation
\citep[see][]{gh02,gh03}.  Thus, ensuring that
the column density values reported here are consistent with both analyses
significantly improves our confidence in these results.

\subsection{Velocities and Doppler Parameters}

     In the global fits described in \S4.1, the ISM absorption lines are
divided into three categories:  atomic lines, H~I lines, and H$_{2}$ lines.
The heliocentric centroid velocities of these lines in the JL~9 fit in
Figure~1 are $-24.2$, $-29.2$, and $-17.2$ km~s$^{-1}$ for the atomic, H~I,
and H$_{2}$ lines, respectively.  These values are of questionable accuracy
due to the uncertainty in the absolute wavelength scale (see \S3), but
these velocities are not too different from the ISM velocity expected for
the Local Interstellar Cloud (LIC) along this line of sight.  The LIC
vector of \citet{rl95} predicts a line of sight velocity
of $-12.6$ km~s$^{-1}$, which is not too far away from the measured
velocities listed above considering that multiple ISM components will
certainly exist along this lengthy line of sight that will shift the
measured line centroids away from the LIC velocity.  Similar rough
agreement is found for LSS~1274, where the predicted LIC velocity is
$0.3$ km~s$^{-1}$ and the measured velocities are $4.2$, $5.9$, and
$9.2$ km~s$^{-1}$ for the atomic, H~I, and H$_{2}$ lines, respectively.

     In the global fits, Doppler parameters of the atomic lines are
computed from $T$ and $\xi$ values, which are free parameters of the fits.
Since the lines are not resolved and we do not know the ISM velocity
structure along the lines of sight, the meaning of these parameters is
very questionable.  In all of the fits, the $b$ values are dominated by
the nonthermal broadening parameter ($\xi$).  For JL~9 we find
$\xi\approx 11$ km~s$^{-1}$ and for LSS~1274 we find
$\xi\approx 6$ km~s$^{-1}$.  The H~I lines also show larger line
widths for the JL~9 line of sight, where we typically find
$b({\rm H~I})\approx 15$ km~s$^{-1}$ for JL~9 compared with
$b({\rm H~I})\approx 13$ km~s$^{-1}$ for LSS~1274.
The larger $\xi$ and $b$ values for the JL~9 line of sight may imply
a broader distribution of ISM velocity components, at least for the atomic
lines.  In contrast, we typically find
$b({\rm H_{2}})\approx 3$ km~s$^{-1}$ for both lines of sight.

\subsection{Column Densities}

     The most important fit parameters are the column densities.
The accuracy of a column density measurement depends on where the set of
available lines are on the curve of growth \cite[see, e.g.,][]{ls78}.
Optically thin lines in the linear part of the curve of growth provide
the best constraints on the column densities, although weak lines are
more sensitive to errors induced by low signal-to-noise, unidentified
blends, and uncertainties in continuum placement.  Excellent constraints
on the column density are also provided by very strong lines with damping
wings, which are in the square-root part of the curve of growth
(i.e., H~I Ly$\beta$ and most of the H$_{2}$ $J=0-1$ lines), although
extapolating an accurate continuum over these very broad lines can be
problematic.  Intermediate saturated lines without damping wings in the
flat part of the curve of growth are not very sensitive diagnostics for
column densities.  Column densities derived solely from these lines can be
very sensitive to certain assumptions involved in any fit; e.g., how
Doppler parameters and instrumental LSFs are treated \citep[see][]{gh02}.
Thus, these column densities are flagged in Table~3 as
being potentially unreliable.  The individual column density measurements
and the lines from which they are derived are now discussed in detail.

\subsubsection{Hydrogen}

     There are a large number of H~I lines in the FUSE spectra (see
Fig.~1), but most are highly saturated and located on the flat part of
the curve of growth.  The exceptions are the two strongest lines,
Ly$\beta$ at 1025.7~\AA\ and Ly$\gamma$ at 972.5~\AA, which are shown in
Figure~2.  Our derivation of precise H~I column densities relies on the
existence of substantial damping wings for these lines, especially for
Ly$\beta$.  The figure shows the spread of fits suggested by the
$\pm 2\sigma$ error bars in the H~I column densities quoted in Table~3.
Figure~2 illustrates the importance of correcting for the strong H$_{2}$
absorption along the blue sides of both the Ly$\beta$ and Ly$\gamma$ lines.
Since Ly$\beta$ and Ly$\gamma$ have damping wings, the Ly$\alpha$ line at
1216~\AA\ surely does as well.  The Ly$\alpha$ line is not accessible to
FUSE, but the {\em International Ultraviolet Explorer} (IUE) observed this
line for JL~9 in 1984.  From these data, \citet{ad94} derived
a value of $\log N({\rm H~I})=20.79\pm 0.14$ for JL~9.  This agrees very
well with our $\log N({\rm H~I})=20.78\pm 0.10$ result from the better
quality FUSE data.  Unfortunately, there are no IUE or HST observations of
the Ly$\alpha$ line of LSS~1274.

\subsubsection{Deuterium}

     Located $-82$ km~s$^{-1}$ from every H~I Lyman line is a D~I Lyman
line.  Figure~3 shows the D~I lines that provide the best constraints
on the D~I column density, and shows the spread of fits suggested by
the $\pm 2\sigma$ error bars in the D~I column densities quoted in Table~3.
All other D~I lines are either saturated and highly blended with H~I, or
they are heavily blended with other strong ISM absorption lines that
happen to lie near them (see Fig.~1).  Even for the Lyman-5 and Lyman-10
lines in Figure~3 there are weak H$_{2}$ blends that must be
taken into account in fitting these lines.  The LSS~1274 D~I measurement
in Table~3, $\log N({\rm D~I})=15.86\pm 0.18$ (2$\sigma$ error), is in
excellent agreement with the $\log N({\rm D~I})=15.87\pm 0.10$ (1$\sigma$
error) result from \citet{gh03} based on only the Owens.f
analysis.

     The H~I, D~I, and O~I lines become increasingly crowded below 930~\AA,
in addition to the ever present H$_{2}$ lines (see Fig.~1).  The Lyman-13
D~I line at 916.2~\AA\ is the highest D~I Lyman line that has ever been
detected, although it has also been observed and measured previously for
Feige~110, HD~195965, and HD~191877 \citep{sdf02,cgh03}.
Figure~1 suggests that this may be the highest D~I Lyman
line that one can {\em ever} hope to clearly detect, since absorption from
H~I, O~I, N~II, and H$_{2}$ will probably always obscure higher D~I lines
at shorter wavelengths.  The D~I Lyman-13 line will become saturated for
column densities of $\log N({\rm D~I})\gtrsim 16.5$, so measuring a
precise D~I column density for lines of sight with higher columns will
be impossible.  Our measured D~I column densities listed in Table~3 are
only about a factor of 4 below this limit.  Thus, our two targets are
near the high column density limit where direct D/H measurements are still
possible.  Beyond this limit, the only way to measure D/H is through
deuterated molecular lines, but interpretation of these lines requires
detailed chemical modeling \citep{dal00}.

\subsubsection{Oxygen}

     A large number of O~I lines exist in our spectra (see Fig.~1), but
all but one are saturated and located on the flat part of the curve of
growth.  The one exception is the O~I line at 974.07~\AA, which is blended
with two H$_{2}$ lines.  Because of the importance of the O~I 974.07~\AA\
line in deriving a precise O~I column density, we try fitting it alone as
well as including it in the global fits.  These fits are shown in Figure~4
for both the SiC2A and SiC1B segments.  Although these fits are not
part of global fits, the results of the global fits are used to constrain
the centroids and Doppler parameters of the O~I and blended H$_{2}$ lines,
and are also used to determine the assumed LSFs.  Note the significantly
lower resolution of the SiC1B data in this wavelength region for both stars.

     For LSS~1274, the O~I column densities suggested by the two fits in
Figure~4 are nicely consistent with the results from the global fits.
However, this is not the case for JL~9, at least for the SiC2A data.
The global JL~9 fits lead to very poor fits to the SiC2A O~I 974.07~\AA\
line.  Even the SiC2A JL~9 fit in Figure~4 does not look very good,
although the combined fit to O~I and the two H$_{2}$ lines has a reasonable
$\chi^{2}_{\nu}=1.10$ value.  The observed
O~I absorption seems to be too redshifted and too blended with the H$_{2}$
absorption.  The reason for this is unclear, but we decide that for JL~9
the results of the O~I fits in Figure~4 lead to better fits and more
believable O~I column densities than the global fits (at least for the
SiC2A data). Thus, the Figure~4 fits are the primary source of the JL~9
O~I column density reported in Table~3, in addition to the Owens.f results.
Note that the LSS~1274 O~I measurement in Table~3,
$\log N({\rm O~I})=17.65\pm 0.15$ (2$\sigma$ error), is in
perfect agreement with the $\log N({\rm O~I})=17.62\pm 0.08$ (1$\sigma$
error) result from \citet{gh03} based on only the Owens.f
analysis.

     Since charge exchange should keep O and H in the same ionization
state in the ISM, O~I has been used as a proxy for H~I in cases where H~I
cannot be measured accurately, and D/O has been used as a proxy for D/H
\citep[see][]{gh03}.  However, in the high column density
regime in which our two lines of sight exist, H~I can be measured more
accurately than O~I (see Table~3).  This is because the high column
densities lead to almost all of the O~I diagnostics being saturated, but
also lead to strong damping wings for H~I Ly$\beta$, which provide
excellent constraints on the H~I column (see \S4.3.1).

\subsubsection{Nitrogen}

     The primary constraints on the N~I column density are the N~I lines
between $950-965$~\AA\ (see Fig.~1).  The LSS~1274 N~I column density
measured from these lines is the only column density in which the
global fits and the Owens.f analysis are not in good agreement.  The
Owens.f analysis focuses only on the optically thin lines, which should
provide the best constraints.  This analysis therefore relies on
the N~I 951.08~\AA, 951.29~\AA, 955.52~\AA, 955.88~\AA, and 959.49~\AA\
lines \citep{gh03}.  However, in the global fits the
absorption features near the 951.29~\AA\ and 959.49~\AA\ lines are seen to
be shifted significantly from the expected positions of the N~I absorption
and are therefore essentially ignored.  The stronger N~I 952.52~\AA\ line,
which is not considered in the Owens.f analysis, seems to suggest lower
column densities than the weaker lines mentioned above.  The global fits
are driven to fit this and other strong N~I lines well instead of fitting
the weak lines well.  The 955.88~\AA\ line is blended with an H$_{2}$ line,
complicating its analysis, and it is only in the SiC1B data that there is a
weak feature that might be N~I $\lambda$955.52 --- it is not apparent in
SiC2A.  In short, in the global fits some (possibly all) weak lines used in
the Owens.f analysis are seen as being contaminated by blends or continuum
variations.  On the other hand, all the strong lines used in the global
fit (except maybe $\lambda$952.52) are saturated, which makes the
column density measurements from the global fits subject to systematic
effects due to the unknown velocity structure of the line of sight and
the unknown shape of the LSF \citep{gh02}. In addition, many
of the strong N~I lines are blended, complicating their analysis.

     The results of the global fit and the Owens.f fits are $\log
N({\rm N~I})=16.30\pm 0.28$ and $16.73\pm 0.10$, respectively.  It is
difficult to be sure which of these two analyses gives a more
reliable answer, so the N~I column density listed in Table~3,
$\log N({\rm N~I})=16.52\pm 0.36$, is a compromise value and the assumed
error bar is expanded to encompass the 1$\sigma$ error ranges suggested
by both analyses.  This leads to a larger uncertainty than one would
expect for an atomic species with a seemingly good selection of
absorption lines.  Nevertheless, this illustrates the wisdom of
considering two independent analyses in deriving our column densities,
since the two approaches can reveal potential problems and systematic
errors that would otherwise be unrecognized.

\subsubsection{Other Atomic Lines}

     The measured column densities in Table~3 yet to be discussed are
those of C~I, P~II, Ar~I, and Fe~II.  The C~I measurements are entirely
based on the C~I 945.19~\AA\ line, which is not saturated and therefore
provides a reasonably accurate C~I column.  There are a large number
of Fe~II lines in the spectra, including many unsaturated ones (see Fig.~1),
but several of these lines are not fit well (e.g., the 926.21~\AA,
926.90~\AA, and 1142.47~\AA\ lines).  The nature of the discrepancy is the
same for both our lines of sight, so we suspect inaccurate oscillator
absorption strengths for these lines, and such inaccuracies could be
significant sources of systematic uncertainty for the Fe~II column
densities.  \citet{jch00} have previously found significant problems
with a few other Fe~II lines in the FUSE bandpass, though not the ones
mentioned above.

     The Ar~I column density is derived from two lines at 1048.22~\AA\ and
1066.66~\AA, and the P~II column density is measured from detected lines at
961.04~\AA, 963.80~\AA, and 1152.82~\AA.  Unfortunately, the Ar~I and P~II
lines are all saturated and located on the flat part of the curve of
growth, meaning that our column density measurements must be considered to
be potentially unreliable, as discussed above, despite our efforts to be
particularly conservative regarding uncertainty estimates.

\subsubsection{Molecular Hydrogen}

     Figure~1 shows that absorption lines of molecular hydrogen are
ubiquitous throughout the FUSE spectra of JL~9, as they are for LSS~1274
as well.  The H$_{2}$ lines actually outnumber the detected
atomic lines.  Nearly all of the numerous H$_{2}$(J=0) and H$_{2}$(J=1)
lines have strong damping wings.  This leads to particularly precise
column density measurements for these two H$_{2}$ levels, which have
smaller error bars than any of our other measurements (see Table~3).
In contrast, the H$_{2}$(J=2) and H$_{2}$(J=3) lines are all saturated and
in the flat part of the curve of growth.  Despite the existence of a large
number of such lines, we still find large error bars for the column
densities of these H$_{2}$ levels.  Error bars are lower for the
H$_{2}$(J=4) and H$_{2}$(J=5) column densities, since there are optically
thin H$_{2}$ lines in the FUSE spectra that provide better constraints.

     We are able to measure column densities for the six lowest H$_{2}$
levels, and in Figure~5 we plot the relative level populations for both the
JL~9 and LSS~1274 lines of sight.  The populations are very similar for
these two cases.  Figure~5 also illustrates thermal populations for
$T=50-200$~K, demonstrating that the observed level populations are not
well represented by a single thermal population.  This is typical for
interstellar H$_{2}$.  The usual explanation for this is that the $J\geq 2$
levels are nonthermally populated by radiative deexcitation from
high levels, which are pumped by UV photons
\citep[e.g.,][]{jhb73,tps00,blr01}.  Despite these nonthermal
effects, the H$_{2}$(J=1)/H$_{2}$(J=0) column density ratio is
generally assumed to be indicative of the actual thermal temperature
of the H$_{2}$ gas.  This ratio suggests temperatures of $T=89\pm 6$~K
and $T=64\pm 5$~K for JL~9 and LSS~1274, respectively.  These values
are consistent with the $T=77\pm 17$~K average ISM H$_{2}$ temperature
found by \citet{bds77} for 61 lines of sight observed by
{\em Copernicus}, and with the average temperature of
$T=68\pm 15$~K measured by \citet{blr02} for 23 lines of
sight observed by FUSE.

     The total H$_{2}$ column densities towards JL~9 and LSS~1274 are
$\log N({\rm H_{2}})=19.25\pm 0.03$ and
$\log N({\rm H_{2}})=19.10\pm 0.04$, respectively.  The hydrogen molecular
fraction can be defined as
$f({\rm H_{2}})=2N({\rm H_{2}})/[2N({\rm H_{2}})+N({\rm H~I})]$.  We find
low molecular fractions of $f({\rm H_{2}})=0.056\pm 0.012$ and
$f({\rm H_{2}})=0.032\pm 0.006$ for the JL~9 and LSS~1274 lines of sight,
respectively.

\section{THE GALACTIC D/H RATIO}

\subsection{A Revised Galactic Gas-Phase D/H Estimate}

     The quantity that we are most interested in measuring for our two
lines of sight is the interstellar gas-phase D/H ratio.  Based on the H~I
and D~I column densities listed in Table~3, we find that
${\rm D/H}=(1.00\pm 0.37)\times 10^{-5}$ for JL~9 and
${\rm D/H}=(0.76\pm 0.36)\times 10^{-5}$ for LSS~1274 (2$\sigma$
uncertainties).  The molecular fraction for these
lines of sight is very low (see \S4.3.6), so these atomic D/H ratios
should be excellent measurements of the gas-phase D/H ratio.

     In order to place these new measurements into their proper context,
we have compiled a large number of D/H measurements for comparison.  These
D/H values are listed in Table~4.  Previous lists of D/H measurements from
\citet{prm92}, \citet{jll98}, and \citet{hwm02} were invaluable
in compiling Table~4, but the references in the table are to the original
sources.  The list is meant to be as comprehensive as possible, but
D/H values with extremely large uncertainties (i.e., greater than a factor
of 2) are not listed, and we only list the most recent D/H measurements for
each line of sight.  Many of the older {\em Copernicus} measurements have
been superceded by better observations and analyses.  {\em Hipparcos}
distances are used when available, but some of the distances are simply
those quoted in the references.  The ``Satellite'' column in Table~4
identifies the source of the D~I measurement, although for G191-B2B and
HZ~43 observations of D~I Ly$\alpha$ from HST were considered in addition
to the FUSE data.

     The error bars on the D/H values in Table~4 are assumed to be
1$\sigma$.  Deciding how to interpret the error bars given in the
literature can be difficult.  Some analyses quote uncertainties as either
1$\sigma$ or 2$\sigma$, but many others do not.  Gaussian statistics are
not necessarily applicable to D/H analyses, depending on the method used,
and as a consequence the quoted uncertainties cannot precisely be given a
``1$\sigma$'' or ``2$\sigma$'' label.  An example is the global fit
approach described in \S4.1, which yields uncertainties than cannot be
easily labeled ``1$\sigma$'' or ``2$\sigma$'', although we have argued that
they can be considered close to ``2$\sigma$''.  (However, note that the
Owens.f analysis also used here {\em does} compute formal Gaussian
statistical uncertainties.)  In any case, we follow McCollough's (1992)
example of generally assuming {\em Copernicus} uncertainties are 1$\sigma$,
but for many of the HST measurements without a clear definition of the
quoted uncertainties, we assume that they are 2$\sigma$.  We will return
to this issue below.  Note that for WD~0621-376 and WD~2211-495 we have
followed the example of \citet{hwm02} in assuming 40\% errors for
$\log N({\rm H~I})$ and D/H.

     In Figure~6, the D/H values are plotted versus distance and H~I
column density.  There appears to be a strong tendency for the D/H ratio
to be low at the longest distances and the highest column densities.
Of the 17 D/H measurements for lines of sight over 100~pc, there are only
two that are greater than the gas-phase Local Bubble value of
${\rm (D/H)_{LBg}}\approx 1.5\times 10^{-5}$ \citep{jll98}, while 15 are
lower.  The apparent dependence of D/H on distance and $N({\rm H~I})$ is
strengthened considerably by the addition of the two new D/H data points
from the JL~9 and LSS~1274 lines of sight, which are both lengthy, high
column density lines of sight with low values of D/H.  Figure~6b has been
divided into three column density regimes:  $\log N({\rm H~I})<19.2$,
$19.2<\log N({\rm H~I})<20.5$, and $\log N({\rm H~I})>20.5$.
The D/H ratio appears constant in the lowest and highest regimes, although
at different values, while D/H appears to be variable in the intermediate
region.  A similar dependence was found by \citet{gh03} for
the D/O and D/N ratios.  The apparent variability of D/H beyond 100~pc has
been noted previously \citep{ebj99,hwm02,gh03,jll03}.
These results can be explained if
the gas-phase D/H in the ISM is variable on some well-defined size scale,
with regions of constant D/H having a typical column density of
$\log N({\rm H~I})\sim 19$ (like the Local Bubble).  However, very long
sight lines sample many of these regions, so the D/H variations should
average out for sufficiently lengthy lines of sight.  Thus, for
$\log N({\rm H~I})\gtrsim 20.5$ we once again start to see roughly constant
D/H.

     We consider the lowest column density regime in Figure~6b to be that
of the Local Bubble, and D/H here is consistent with a constant value of
${\rm (D/H)_{LBg}}=(1.56\pm 0.16)\times 10^{-5}$ (1$\sigma$ standard
deviation).  The 21 lines of sight from which this ${\rm (D/H)_{LBg}}$ value
is computed are flagged with ``LBg'' in the second-to-last column of Table~4.
It is worth noting that three out of the 21 LBg D/H measurements have error
bars that do not overlap the average ${\rm (D/H)_{LBg}}$ value.  This is a
reasonable fraction for 1$\sigma$ error bars, which is an
{\em a posteriori} argument for the quoted error bars in Table~4 being
reasonably accurate.  If the error bars were increased significantly, the
agreement with the average ${\rm (D/H)_{LBg}}$ value would look {\em too}
good.  Given the apparent constancy of ${\rm (D/H)_{LBg}}$, it is
appropriate to replace the standard deviation error with a standard
deviation of the mean, so our final value for the Local Bubble gas-phase
D/H is ${\rm (D/H)_{LBg}}=(1.56\pm 0.04)\times 10^{-5}$.

     The constancy of both D/H and D/O within the Local Bubble is known
from previous work, and our new ${\rm (D/H)_{LBg}}$ value is in good
agreement with previous Local Bubble measurements
\citep{jll98,hwm02,gh03,jll03}.  This
homogeneity of local gas-phase D/H values likely results from gas within
the Local Bubble being well mixed, with a common history initiated by
supernovae events and O~star winds emerging from the Scorpio-Centaurus
Association a few million years ago.  The D/H values within the Local
Bubble are indicative of conditions in its warm clouds, which have similar
ages, chemical compositions, and are bathed in similar UV radiation fields.
Note that \citet{gh03} derive a value of
${\rm (D/H)_{LBg}}=(1.32\pm 0.08)\times 10^{-5}$ from D/O and a typical ISM
O/H ratio, and they discuss possible reasons why this value does not agree
precisely with the direct D/H measurements.

     The high column density boundary where D/H becomes roughly constant
again is somewhat unclear, but in Figure~6b we draw it at
$\log N({\rm H~I})=20.5$, corresponding to a distance of $d\approx 500$~pc.
There are four D/H measurements with larger columns, which are flagged with
``LDg'' in Table~4.  All of these measurements are from FUSE (including the
two new ones presented here).  We believe that these long lines of sight
provide the best measurements of the local-disk gas-phase D/H ratio,
${\rm (D/H)_{LDg}}$, since they sample far more regions of the ISM than do
shorter lines of sight with lower columns.  However, with only four
measurements we cannot rule out the possibility that other Galactic lines
of sight with similarly high columns might
exist that yield significantly different D/H.  The Feige~110
[$\log N({\rm H~I})=20.14^{+0.07}_{-0.10}$] and $\gamma^{2}$~Vel
[$\log N({\rm H~I})=19.710\pm 0.026$] lines of sight illustrate this
possibility clearly, as both have D/H twice as high as that seen for the
four $\log N({\rm H~I})>20.5$ sight lines.  The $\alpha$~Cru D/H value is
also high, but with very large error bars.  Despite this, the tendency for
high column lines of sight to yield low D/H seems substantial, considering
that 14 of the 17 measurements with $\log N({\rm H~I})>19.2$ have
${\rm D/H}<{\rm (D/H)_{LBg}}$.

     Collectively, the four $\log N({\rm H~I})>20.5$ lines of
sight suggest ${\rm (D/H)_{LDg}}=(0.85\pm 0.10)\times 10^{-5}$ (1$\sigma$
standard deviation).  Since the four D/H measurements are individually
consistent with this value, we can assume constancy and replace the
1$\sigma$ error quoted above with a standard deviation of the mean, as we
did for ${\rm (D/H)_{LBg}}$, thereby obtaining a final value of
${\rm (D/H)_{LDg}}=(0.85\pm 0.09)\times 10^{-5}$.  This can be compared to
the ${\rm (D/H)_{LDg}}=(0.52\pm 0.09)\times 10^{-5}$ result found by
\citet{gh03} for long distances from D/O measurements
combined with O/H from \citet{dmm98}, and it agrees even
better with their ${\rm (D/H)_{LDg}}=(0.86\pm 0.13)\times 10^{-5}$ result
similarly derived from D/N measurements and N/H from \citet{dmm97}.

     In the past, the well determined Local Bubble D/H value,
${\rm (D/H)_{LBg}}$, has been assumed to be characteristic of the Galaxy as
a whole, but D/H measurements for long lines of sight suggest that
${\rm (D/H)_{LBg}}$ is {\em not} representative of the Galactic
ISM, and that the Local Bubble D/H value is actually a factor of 2 higher
than the true average gas-phase local-disk D/H value.
However, is the total (i.e., gas plus dust) local-disk D/H ratio
[${\rm (D/H)_{LD}}$] equal to the gas-phase value [${\rm (D/H)_{LDg}}$], or
could the Local Bubble value [${\rm (D/H)_{LBg}}$] actually be a better
estimate for ${\rm (D/H)_{LD}}$?  The answer depends on the cause of the
D/H variability seen in Figure~6, which we now discuss in some detail.

\subsection{What is the Cause of the D/H Variability?}

\subsubsection{Variable Astration}

     Possible causes for the D/H variability have been previously
discussed by \citet{ml99} and \citet{hwm02}.
One explanation for the D/H variability apparent in
Figure~6 is that the ISM is simply not well mixed on distance scales of
a few hundred parsecs and column density scales of
$19.2<\log N({\rm H~I})<20.5$, despite being relatively homogeneous on
smaller and larger scales.  If this is the case, different ISM regions may
be characterized by different amounts of stellar processing (i.e.,
astration) and this could therefore explain the observed D/H variations.
Supernovae are the primary drivers of ISM mixing on large distance scales,
although paradoxically, they are also potential sources of abundance
inhomogeneity.  The degree of mixing in the Galactic ISM has been
studied by hydrodynamic models of the ISM \citep[e.g.,][]{maa02}.
For the current Galactic supernova rate, it is found that mixing
time scales are of order 350~Myr.  Thus, the local ISM would not have a
constant D/H if sources of inhomogeneities, mainly supernovae, have
occurred within this timescale.  The conclusion is that inhomogeneities
can potentially exist in the local ISM, though their existence is by no
means certain.

     Arguments against this have been made based on measurements of O/H,
which find no significant spatial variations \citep*{dmm98,mka03}.
However, these particular analyses only sample
lines of sight with high column densities of $20.15<\log N({\rm H~I})<21.5$,
and based on Figure~6b we are now arguing that D/H may not vary in this
high column density regime either.  It is only at shorter distance scales
with smaller columns that variability is apparent.

     In the variable astration scenario, the long distance, high-column
${\rm (D/H)_{LDg}}=(0.85\pm 0.09)\times 10^{-5}$ value derived above would
provide the best estimate for ${\rm (D/H)_{LD}}$.  This suggests
a significantly greater degree of astration than has been previously
assumed.  When compared with ${\rm (D/H)_{prim}}$ (see \S1), our
${\rm (D/H)_{LDg}}$ value implies a deuterium destruction factor of
$3.3\pm 0.6$.  This higher destruction factor might be a problem for
Galactic chemical evolution models, since most models can account for
destruction factors of only $1.5-3$ \citep{np96,mt98,cc02}.
However, there are nonstandard models involving prominent Galactic winds
that lead to significantly higher destruction factors
\citep[e.g.,][]{evf95,ss97}.

\subsubsection{Depletion of Deuterium onto Dust Grains}

     We can assume ${\rm (D/H)_{LD}}={\rm (D/H)_{LDg}}$ only if deuterium
is not significantly depleted onto dust preferentially to hydrogen.
However, it has been argued that D {\em can} be preferentially depleted in
this manner.  \citet{mj82} first suggested that cold interstellar grains
could remove a significant amount of deuterium from the gas phase.
\citet{btd03} recently showed that extreme enrichments of deuterium in
carbonaceous grains in the diffuse interstellar medium is thermodynamically
favored.  The zero-point energy of the C-D bond exceeds that of the C-H
bond by 0.092~eV, whereas the zero-point energy of the H-D bond exceeds
that of the H-H bond by 0.035~eV.  Thus, in thermodynamic equilibrium
carbonaceous dust grains can have deuterium enrichments by a
factor $> 10^4$ for grain temperatures $T_{\rm grain} < 70$~K.  Large
polycyclic aromatic hydrocarbon (PAH) molecules can also be highly
deuterium enriched if they are as cold as the grains \citep{ep04}.
Under these conditions, the abundance of deuterium in the gas phase of the
ISM would be reduced by $1\times 10^{-5}$, which is sufficient to explain
the low values of D/H for the lines of sight with large column
densities.

     Is this model of variable depletion of deuterium from the gas phase
realistic given the dynamics and radiation environment of the ISM?  Cold
interstellar gas in clouds is known to have molecules with deuterium
enrichment by factors of $10^4$ or larger \citep[e.g.,][]{ab03}.
It is unlikely that the lines of sight to JL~9 and LSS~1274 traverse very
much cold gas since the molecular hydrogen fractions are 0.056 and
0.032, respectively (see \S4.3.6).  It is interesting, however, that the
temperatures of the H$_2$ gas for these lines of sight are $89\pm 6$~K
and $64\pm 5$~K, sufficiently low for highly deuterated molecules to form.
A more relevant consideration is the presence of dust grains even in warm
gas, as indicated by metal depletions for short lines of sight in the Local
Bubble.  Given the low gas density of the diffuse ISM, cold grains
coexist with warm ($T\approx 7000$~K) gas.

     These considerations lead to the following possible explanation for
the three D/H regimes shown in Figure~6b.  Strong shocks produced by
supernovae and the winds of hot stars evaporate grains, forcing all of the
matter into the gas phase.  Over time this gas cools, forming grains that
as they cool preferentially remove deuterium from the gas phase.  Lines of
sight traversing regions that were recently shocked should have most or all
of the material in the gas phase and thus the highest D/H ratio.  The Local
Bubble is one such region.  Gas-phase D/H in the Local Bubble is likely
close to ${\rm (D/H)_{LD}}$, since there are only a few lines of
sight (e.g., $\alpha$~Cru, Feige 110, and $\gamma^{2}$~Vel) with slightly
higher gas-phase D/H values.  The unusual, high D/H values for these few
lines of sight may indicate that they pass through regions that were
shocked more recently than the Local Bubble, or considering the small
number of these high D/H results, they could just
be statistical anomalies.  Lines of sight with high
column densities [$\log N(H~I) > 20.5$] likely traverse a statistical
average of the Galactic disk material, consisting mostly of gas that has not
been shocked for a long time and thus with low gas-phase D/H.  The
intermediate regime with $19.2 < \log N(H~I) < 20.5$, containing a wide
variety of gas-phase values of D/H, does not contain a reasonable
statistical average of shocked and unshocked gas along the individual lines
of sight.

     In this scenario, the Local Bubble value of
${\rm (D/H)_{LBg}}=(1.56 \pm 0.04) \times 10^{-5}$ is a better
estimate for ${\rm (D/H)_{LD}}$ than the global gas-phase
${\rm (D/H)_{LDg}}$ value computed from high column lines of sight in \S5.1.
However, the existence of a few lines of sight with
${\rm D/H}>2\times 10^{-5}$ suggests that deuterium might even be slightly
depleted within the Local Bubble, so if dust depletion is the cause of
the D/H variations we can only really say
${\rm (D/H)_{LD}}\gtrsim {\rm (D/H)_{LBg}}$.  Perhaps future studies of
interstellar dust grains collected within the solar system could provide a
direct determination of the degree of deuterium depletion in the Local Bubble
\citep{pcf99}.  In any case, the deuterium
destruction factor of $1.8\pm 0.3$ suggested
by the Local Bubble D/H value is much easier for Galactic chemical
evolution models to explain than the higher destruction factor of
$3.3\pm 0.6$ implied by the variable astration scenario in \S5.2.1.

\section{Summary}

     We have analyzed FUSE observations of interstellar absorption
for two long lines of sight towards the sdO stars JL~9 and
LSS~1274.  Our results are summarized as follows:
\begin{description}
\item[1.] Using two separate measurement techniques, we have measured
  column densities for many different atomic species and for
  the $J=0-5$ rotational levels of H$_{2}$.  The results are listed in
  Table~3.
\item[2.] We find low molecular fractions of
  $f({\rm H_{2}})=0.056\pm 0.012$ and $f({\rm H_{2}})=0.032\pm 0.006$
  towards JL~9 and LSS~1274, respectively.  The H$_{2}$ gas along
  these lines of sight is found to have a temperature of $T=89\pm 6$~K for
  JL~9 and $T=64\pm 5$~K for LSS~1274, both very typical values for H$_{2}$
  in the Galaxy.
\item[3.] The D/H ratios for the two lines of sight are
  ${\rm D/H}=(1.00\pm 0.37)\times 10^{-5}$ for JL~9 and
  ${\rm D/H}=(0.76\pm 0.36)\times 10^{-5}$ for LSS~1274 (2$\sigma$
  uncertainties).  These D/H values are low compared to the Local Bubble
  value.  When considered with other measurements, these new results
  provide additional crucial evidence that long lines of sight with high
  column densities tend to have low gas-phase deuterium abundances.
  This confirms the results of \citet{gh03}
  based on D/O and D/N measurements, and older published D/H values.
\item[4.] We consider our two new D/H measurements in combination with
  previous Galactic D/H measurements from {\em Copernicus}, HST, IMAPS,
  and FUSE.  Collectively, these data suggest that D/H is constant for
  both low column density [$\log N({\rm H~I})<19.2$] and high column
  density [$\log N({\rm H~I})>20.5$] lines of sight, but variable for
  intermediate columns [$19.2<\log N({\rm H~I})<20.5$].  This suggests
  that regions of constant D/H in the ISM have typical column densities of
  $\log N({\rm H~I})\sim 19$.  However, no variability is seen for
  lines of sight with $\log N({\rm H~I})>20.5$, perhaps due to the
  sampling of a large number of these regions for such long sight lines.
\item[5.] The low column density regime [$\log N({\rm H~I})<19.2$]
  represents the Local Bubble.  Our gas-phase Local Bubble D/H value of
  ${\rm (D/H)_{LBg}}=(1.56 \pm 0.04) \times 10^{-5}$ is consistent with
  previous measurements \citep{jll98,hwm02,jll03}.
  The apparent constancy of D/H at high column densities relies heavily on
  the two new D/H values for JL~9 and LSS~1274.  Since longer, higher
  column lines of sight sample more regions of the ISM, we argue that
  the D/H values for these longer lines of sight provide better estimates
  of the true gas-phase local-disk D/H ratio [${\rm (D/H)_{LDg}}$] than
  the Local Bubble measurements.  For the lines of sight with
  $\log N({\rm H~I})>20.5$ we find
  ${\rm (D/H)_{LDg}}=(0.85\pm 0.09)\times 10^{-5}$.  This is a factor of 2
  lower than ${\rm (D/H)_{LBg}}$, but is similar to the value derived by
  \citet{gh03} from D/O and D/N measurements of lengthy
  sight lines.  (Note that errors quoted above for ${\rm (D/H)_{LBg}}$ and
  ${\rm (D/H)_{LDg}}$ are 1$\sigma$ standard deviations of the mean.)
\item[6.] The cause of the observed D/H variability within the Galaxy
  is uncertain.  We discuss two possible explanations:  1. Variable
  astration and incomplete mixing in the ISM, and 2. Depletion of
  deuterium onto dust grains.  If \#2 is correct, then the low D/H values
  measured for long lines of sight are due to depletion.  In this case,
  the total (i.e., gas plus dust) D/H ratio of the local disk,
  ${\rm (D/H)_{LD}}$, is at least as high as the Local Bubble value of
  ${\rm (D/H)_{LBg}}=(1.56 \pm 0.04) \times 10^{-5}$ instead of being at the
  low gas-phase value of ${\rm (D/H)_{LDg}}=(0.85\pm 0.09)\times 10^{-5}$.
  However, if \#1 is correct, then ${\rm (D/H)_{LD}}={\rm (D/H)_{LDg}}$.
  Deuterium destruction factors can be computed by comparing the Galactic
  and primordial D/H ratios.  Scenario \#1 suggests a destruction
  factor of $3.3\pm 0.6$, while scenario \#2 suggests a value of
  $1.8\pm 0.3$.  The latter is more consistent with the predictions of
  Galactic chemical evolution models.
\end{description}

%[SHOULD I OUTLINE THE STRENGTHS OF THE GLOBAL FITTING APPROACH SOMEWHERE?]

%[SHOULD I MENTION SECONDARY WAVELENGTH CALIBRATION EXERCISE?]

\acknowledgments

This work is based on data obtained for the Guaranteed Time Team by the
NASA-CNES-CSA FUSE mission operated by the Johns Hopkins University.
Financial support to U.\ S.\ participants has been provided by NASA
contract NAS5-32985.  This research has made use of the SIMBAD database,
operated at CDS, Strasbourg, France.  G.\ H.\ was supported by CNES.
This work used the profile fitting procedure Owens.f developed by M.\
Lemoine and the French FUSE Team.  G.\ H.\ would like to thank
B.\ Godard for his help in data processing.

\clearpage

\begin{deluxetable}{lcc}
\tablecaption{Target Star Properties}
\tablecolumns{3}
\tablewidth{0pt}
\tablehead{
  \colhead{Property} & \colhead{JL 9} & \colhead{LSS 1274}}
\startdata
Spectral Type      & sdO          & sdO  \\
RA (2000)          & 19:08:21     &  9:18:56 \\
DEC (2000)         & $-72^{\circ}30^{\prime}34^{\prime\prime}$ &
                     $-57^{\circ}4^{\prime}38^{\prime\prime}$ \\
Gal.\ long.\ (deg) & $322.6$      & $277.0$ \\
Gal.\ lat.\ (deg)  & $-27.0$      &  $-5.3$ \\
V                  & 13.2         & 12.9 \\
B--V               & $-0.28$      & $-0.45$ \\
Distance (pc)      & $590\pm 160$ & $580\pm 100$ \\
\enddata
\end{deluxetable}

\begin{deluxetable}{lclccc}
\tablecaption{FUSE Observations}
\tablecolumns{6}
\tablewidth{0pt}
\tablehead{
  \colhead{Star} & \colhead{Observation ID} & \colhead{Date} &
    \colhead{Aperture} & \colhead{\# of} & \colhead{Exp.\ Time} \\
  \colhead{} & \colhead{} & \colhead{} & \colhead{} &
    \colhead{Exposures} & \colhead{(ksec)}}
\startdata
JL 9     & P3021201 & 2003 May 28   & LWRS &  1 & 16.6 \\
LSS 1274 & P2051702 & 2002 March 8  & MDRS & 11 &  8.0 \\
LSS 1274 & P2051701 & 2002 March 11 & MDRS & 12 & 14.0 \\
LSS 1274 & P2051703 & 2002 May 2    & MDRS & 54 & 63.0 \\
\enddata
\end{deluxetable}

\begin{deluxetable}{lccc}
\tablecaption{Measured Column Densities}
\tablecolumns{4}
\tablewidth{0pt}
\tablehead{
  \colhead{Species} & \colhead{} &
    \multicolumn{2}{c}{$\log N$ (cm$^{-2}$)\tablenotemark{a}} \\
  \colhead{} & \colhead{} & \colhead{JL 9} & \colhead{LSS 1274}}
\startdata
H I          & & $20.78\pm 0.10$ & $20.98\pm 0.08$ \\
D I          & & $15.78\pm 0.12$ & $15.86\pm 0.18$ \\
C I          & & $13.49\pm 0.16$ & $13.55\pm 0.16$ \\
N I          & & $16.27\pm 0.20$ & $16.52\pm 0.36$ \\
O I          & & $17.50\pm 0.33$ & $17.65\pm 0.15$ \\
P II         & & $13.67\pm 0.20$\tablenotemark{b} &
  $13.66\pm 0.23$\tablenotemark{b} \\
Ar I         & & $14.48\pm 0.20$\tablenotemark{b} &
  $14.49\pm 0.35$\tablenotemark{b} \\
Fe II        & & $14.69\pm 0.17$ & $14.81\pm 0.15$ \\
H$_{2}$(J=0) & & $18.87\pm 0.04$ & $18.88\pm 0.05$ \\
H$_{2}$(J=1) & & $18.99\pm 0.04$ & $18.67\pm 0.07$ \\
H$_{2}$(J=2) & & $17.56\pm 0.31$\tablenotemark{b} &
  $17.32\pm 0.50$\tablenotemark{b} \\
H$_{2}$(J=3) & & $17.37\pm 0.43$\tablenotemark{b} &
  $16.85\pm 0.83$\tablenotemark{b} \\
H$_{2}$(J=4) & & $14.76\pm 0.14$ & $14.54\pm 0.16$ \\
H$_{2}$(J=5) & & $14.02\pm 0.27$ & $13.83\pm 0.29$ \\
H$_{2}$(total)& &$19.25\pm 0.03$ & $19.10\pm 0.04$ \\
\enddata
\tablenotetext{a}{With 2$\sigma$ uncertainties.}
\tablenotetext{b}{Potentially unreliable due to measurement solely from
  saturated lines in the flat part of the curve of growth.}
\end{deluxetable}

\begin{deluxetable}{lcccccccc}
\tabletypesize{\scriptsize}
\tablecaption{Compilation of D/H Measurements}
\tablecolumns{9}
\tablewidth{0pt}
\tablehead{
  \colhead{Target} & \colhead{l} & \colhead{b} & \colhead{d} &
    \colhead{$\log N({\rm H~I})$\tablenotemark{a}} &
    \colhead{D/H\tablenotemark{a}} & \colhead{Satellite} &
    \colhead{Flag\tablenotemark{b}} & \colhead{Refs.}\\
  \colhead{} & \colhead{(deg)} & \colhead{(deg)} & \colhead{(pc)} &
    \colhead{} & \colhead{($10^{-5}$)} & \colhead{} & \colhead{} &\colhead{}}
\startdata
$\epsilon$ Eri & 227 &$-48$&$3.218\pm 0.009$ & $17.880\pm 0.035$ &
  $1.4\pm 0.2$         &  HST & LBg& 1\\
Procyon        & 214 &  13 &$3.50\pm 0.01$   & $18.06\pm 0.05$ &
  $1.6\pm 0.2$         &  HST & LBg& 2\\
$\epsilon$ Ind & 336 &$-48$&$3.626\pm 0.009$ & $18.00\pm 0.05$ &
  $1.6\pm 0.2$         &  HST & LBg& 3\\
36 Oph         & 358 &   7 & $5.99\pm 0.04$  & $17.850\pm 0.075$ &
  $1.50\pm 0.25$       &  HST & LBg& 4\\
$\beta$ Gem    & 192 &  23 & $10.34\pm 0.09$ & $18.261\pm 0.037$ &
  $1.47\pm 0.20$       &  HST & LBg& 1\\
Capella        & 163 &   5 & $12.9\pm 0.1$   & $18.239\pm 0.035$ &
 $1.60^{+0.14}_{-0.19}$&  HST & LBg& 2\\
$\beta$ Cas    & 118 & $-3$& $16.7\pm 0.1$   & $18.130\pm 0.025$ &
  $1.70\pm 0.15$       &  HST & LBg& 1\\
$\alpha$ Tri   & 139 &$-31$& $19.7\pm 0.3$   & $18.327\pm 0.035$ &
  $1.32\pm 0.30$       &  HST & LBg& 1\\
$\lambda$ And  & 110 &$-15$& $25.8\pm 0.5$   & $18.45\pm 0.075$ &
  $1.70\pm 0.25$       &  HST & LBg& 3\\
$\beta$ Cet    & 111 &$-81$& $29.4\pm 0.7$   & $18.36\pm 0.05$ &
  $2.20\pm 0.55$       &  HST & LBg& 5\\
HR 1099        & 185 &$-41$& $29.0\pm 0.7$   & $18.131\pm 0.020$ &
  $1.46\pm 0.09$       &  HST & LBg& 5\\
$\sigma$ Gem   & 191 &  23 & $37\pm 1$       & $18.201\pm 0.037$ &
  $1.36\pm 0.20$       &  HST & LBg& 1\\
WD 1634-573    & 330 & $-7$& $37\pm 3$       & $18.85\pm 0.06$ &
  $1.60\pm 0.25$       & FUSE & LBg& 6\\
WD 2211-495    & 346 &$-53$& $53\pm 16$      & $18.76\pm 0.15$ &
  $1.51\pm 0.60$       & FUSE & LBg& 7\\
HZ 43          &  54 &  84 & $68\pm 13$      & $17.93\pm 0.03$ &
  $1.66\pm 0.14$       & FUSE & LBg& 8\\
G191-B2B       & 156 &   7 & $69\pm 15$      & $18.18\pm 0.09$ &
  $1.66\pm 0.45$       & FUSE & LBg& 9\\
WD 0621-376    & 245 &$-21$& $78\pm 23$      & $18.70\pm 0.15$ &
  $1.41\pm 0.56$       & FUSE & LBg&10\\
GD 246         &  87 &$-45$& $79\pm 24$      & $19.110\pm 0.025$ &
 $1.51^{+0.20}_{-0.17}$& FUSE & LBg&11\\
$\alpha$ Vir   & 316 &  51 & $80\pm 6$       & $19.00\pm 0.10$ &
  $1.6^{+1.6}_{-0.6}$  & Copernicus&LBg&12\\
31 Com         & 115 &  89 & $94\pm 8$       & $17.88\pm 0.03$ &
  $2.0\pm 0.2$         &  HST & LBg& 1\\
$\alpha$ Cru   & 300 &   0 & $98\pm 6$       & $19.60\pm 0.10$ &
  $2.5^{+0.7}_{-0.9}$  & Copernicus&...&12\\
BD+28$^{\circ}$4211&82&$-19$&$104\pm 18$     & $19.85\pm 0.02$ &
  $1.39\pm 0.10$       & FUSE &...&13\\
$\theta$ Car   & 290 & $-5$& $135\pm 9$      & $20.28\pm 0.10$ &
  $0.50\pm 0.16$        & Copernicus&...&14\\
$\beta$ CMa    & 226 &$-14$& $153\pm 15$     & $18.20\pm 0.16$ &
  $1.2^{+1.1}_{-0.5}$  & Copernicus&LBg&15\\
$\beta$ Cen    & 312 &   1 & $161\pm 15$     & $19.54\pm 0.05$ &
 $1.26^{+1.25}_{-0.45}$& Copernicus&...&12\\
Feige 110      &  74 &$-59$&$179^{+265}_{-67}$&$20.14^{+0.07}_{-0.10}$ &
  $2.14\pm 0.41$       & FUSE &...&16\\
$\gamma$ Cas   & 124 & $-2$& $188\pm 20$     & $20.04\pm 0.04$ &
  $1.12\pm 0.25$       & Copernicus&...&17\\
$\lambda$ Sco  & 352 & $-2$& $216\pm 42$     & $19.28\pm0.03$ &
  $0.76\pm 0.25$       & Copernicus&...&18\\
$\gamma^{2}$ Vel   & 263 & $-8$& $258\pm 35$     & $19.710\pm 0.026$ &
 $2.18^{+0.22}_{-0.19}$& IMAPS&...&19\\
$\delta$ Ori   & 204 &$-18$& $281\pm 65$     & $20.193\pm 0.025$ &
 $0.74^{+0.12}_{-0.09}$& IMAPS&...&20\\
$\mu$ Col      & 237 &$-27$& $397\pm 87$     & $19.90\pm 0.10$ &
 $0.63^{+1.00}_{-0.23}$& Copernicus&...&12\\
$\iota$ Ori    & 210 &$-20$& $407\pm 127$    & $20.16\pm 0.10$ &
  $1.4^{+0.5}_{-1.0}$  & Copernicus&...&21\\
$\epsilon$ Ori & 205 &$-17$& $412\pm 154$    & $20.40\pm 0.08$ &
  $0.65\pm 0.30$       & Copernicus&...&21\\
$\zeta$ Pup    & 256 & $-5$& $429\pm 94$     & $19.963\pm 0.026$ &
 $1.42^{+0.15}_{-0.14}$& IMAPS&...&19\\
LSS 1274       & 277 & $-5$& $580\pm 100$    & $20.98\pm 0.04$ &
  $0.76\pm 0.18$       & FUSE & LDg &22\\
JL 9           & 323 &$-27$& $590\pm 160$    & $20.78\pm 0.05$ &
  $1.00\pm 0.19$       & FUSE & LDg &22\\
HD 195965      &  86 &   5 & $794\pm 200$    & $20.950\pm 0.025$ &
 $0.85^{+0.17}_{-0.12}$& FUSE & LDg &23\\
HD 191877      &  62 & $-6$&$2200\pm 550$    & $21.05\pm 0.05$ &
 $0.78^{+0.26}_{-0.13}$& FUSE & LDg &23\\
\enddata
\tablenotetext{a}{Quoted uncertainties assumed to be 1$\sigma$ (see text).}
\tablenotetext{b}{Indicates which lines of sight are used to compute the
  gas-phase Local Bubble (LBg) and gas-phase local disk (LDg) D/H values
  described in the text.}
\tablerefs{(1) Dring et al.\ 1997. (2) Linsky et al.\ 1995. (3) Wood
  et al.\ 1996. (4) Wood et al.\ 2000. (5) Piskunov et al.\ 1997. (6) Wood
  et al.\ 2002. (7) H\'{e}brard et al.\ 2002. (8) Kruk et al.\ 2002.
  (9) Lemoine et al.\ 2002. (10) Lehner et al.\ 2002. (11) Oliveira
  et al.\ 2003. (12) York \& Rogerson 1976. (13) Sonneborn et al.\ 2002.
  (14) Allen et al.\ 1992. (15) Gry et al.\ 1985. (16) Friedman et al.\ 2002.
  (17) Ferlet et al.\ 1980. (18) York 1983. (19) Sonneborn et al.\ 2000.
  (20) Jenkins et al.\ 1999. (21) Laurent et al.\ 1979. (22) This paper.
  (23) Hoopes et al.\ 2003.}
\end{deluxetable}

\clearpage

\begin{figure}
\plotfiddle{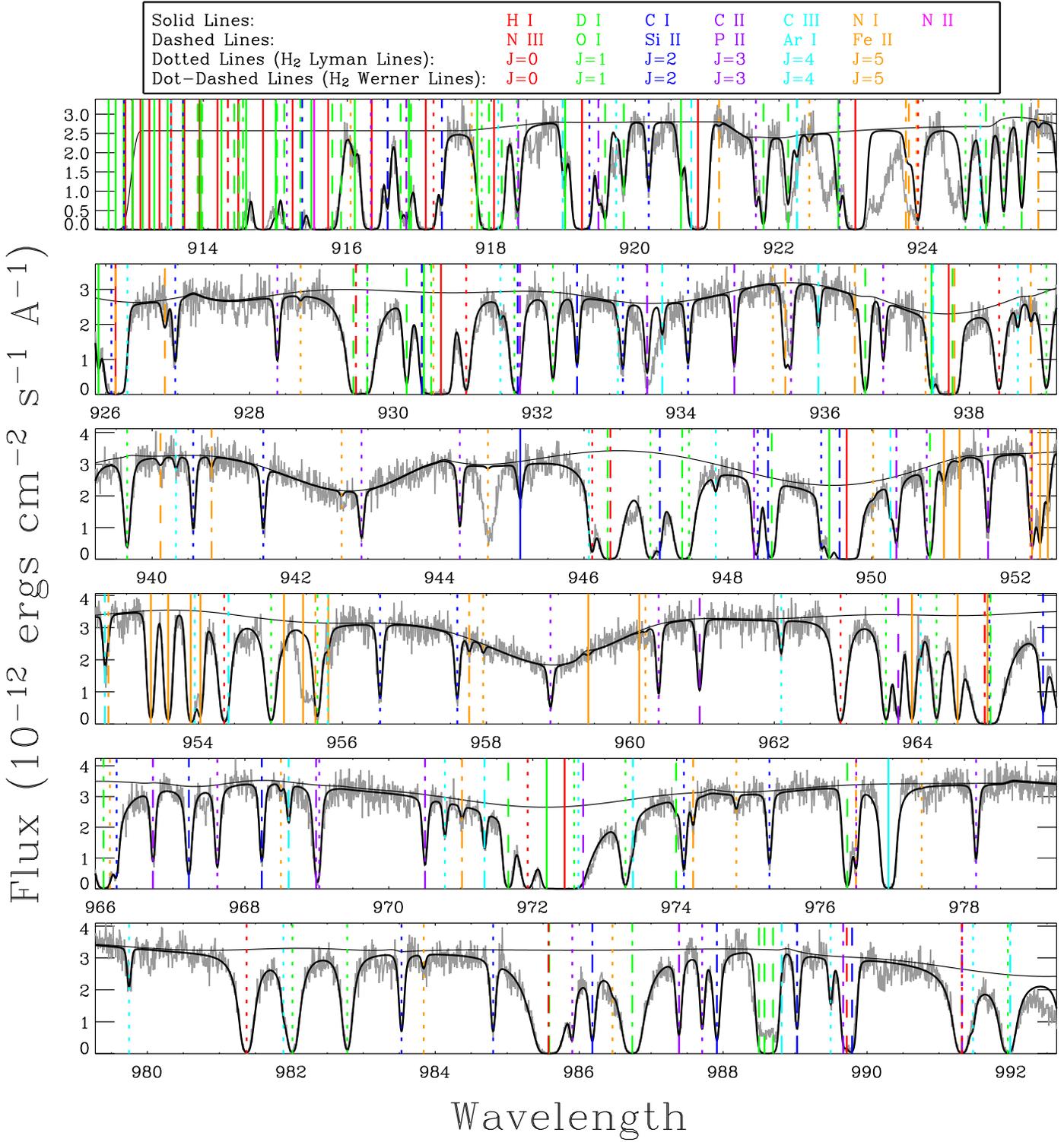}{7.5in}{0}{90}{90}{-390}{0}
\caption{The FUSE spectrum of JL~9, and a fit to the ISM absorption lines
  seen in the spectrum.  Vertical lines of various types and colors
  indicate the locations of ISM absorption lines included in the fit,
  where a key at the top of the figure identifies the lines.  The H$_{2}$
  lines are separated into Lyman band lines and Werner band lines.}
\end{figure}

\clearpage

\setcounter{figure}{0}
\begin{figure}
\plotfiddle{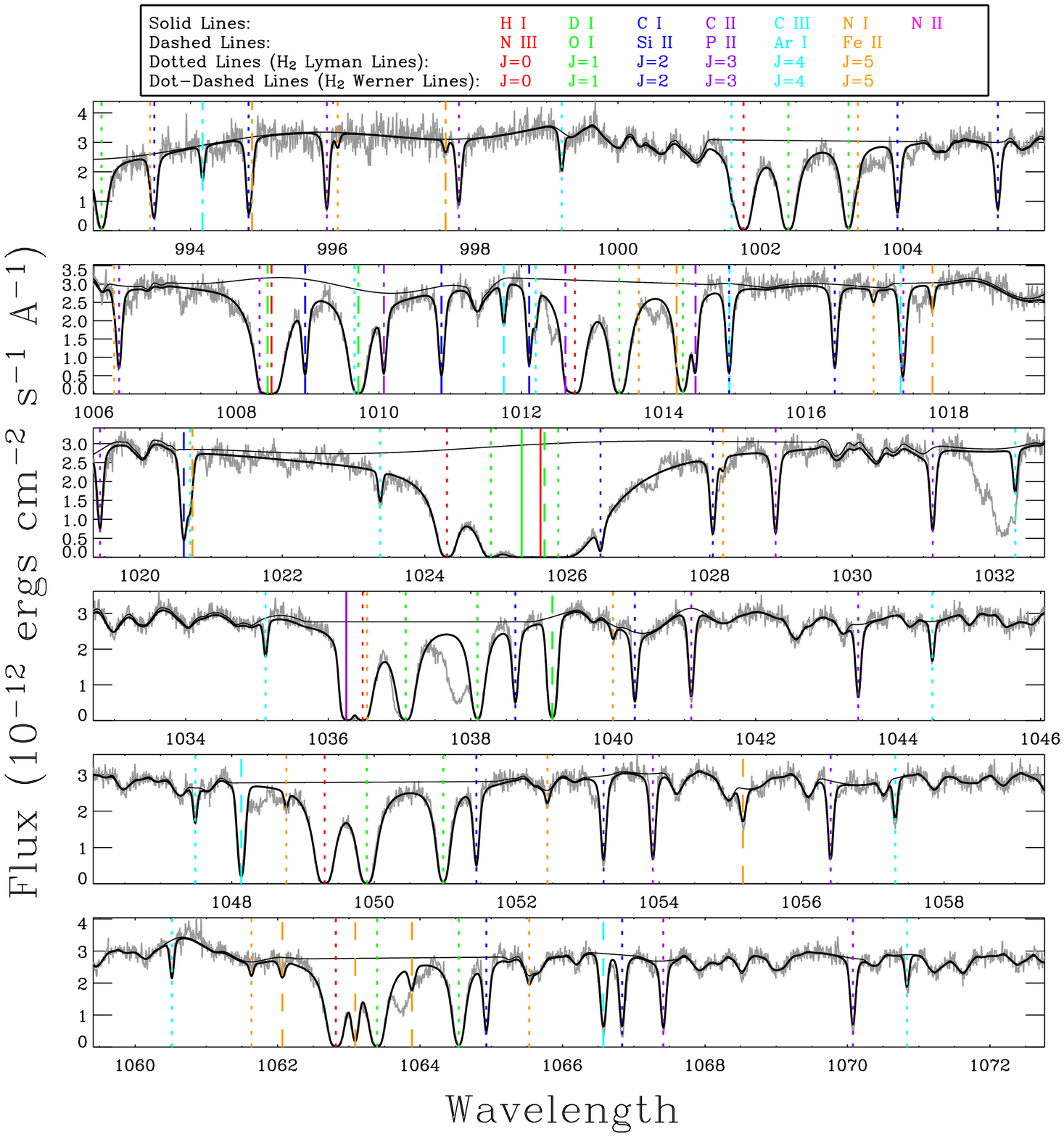}{7.5in}{0}{90}{90}{-390}{0}
\caption{(continued)}
\end{figure}

\clearpage

\setcounter{figure}{0}
\begin{figure}
\plotfiddle{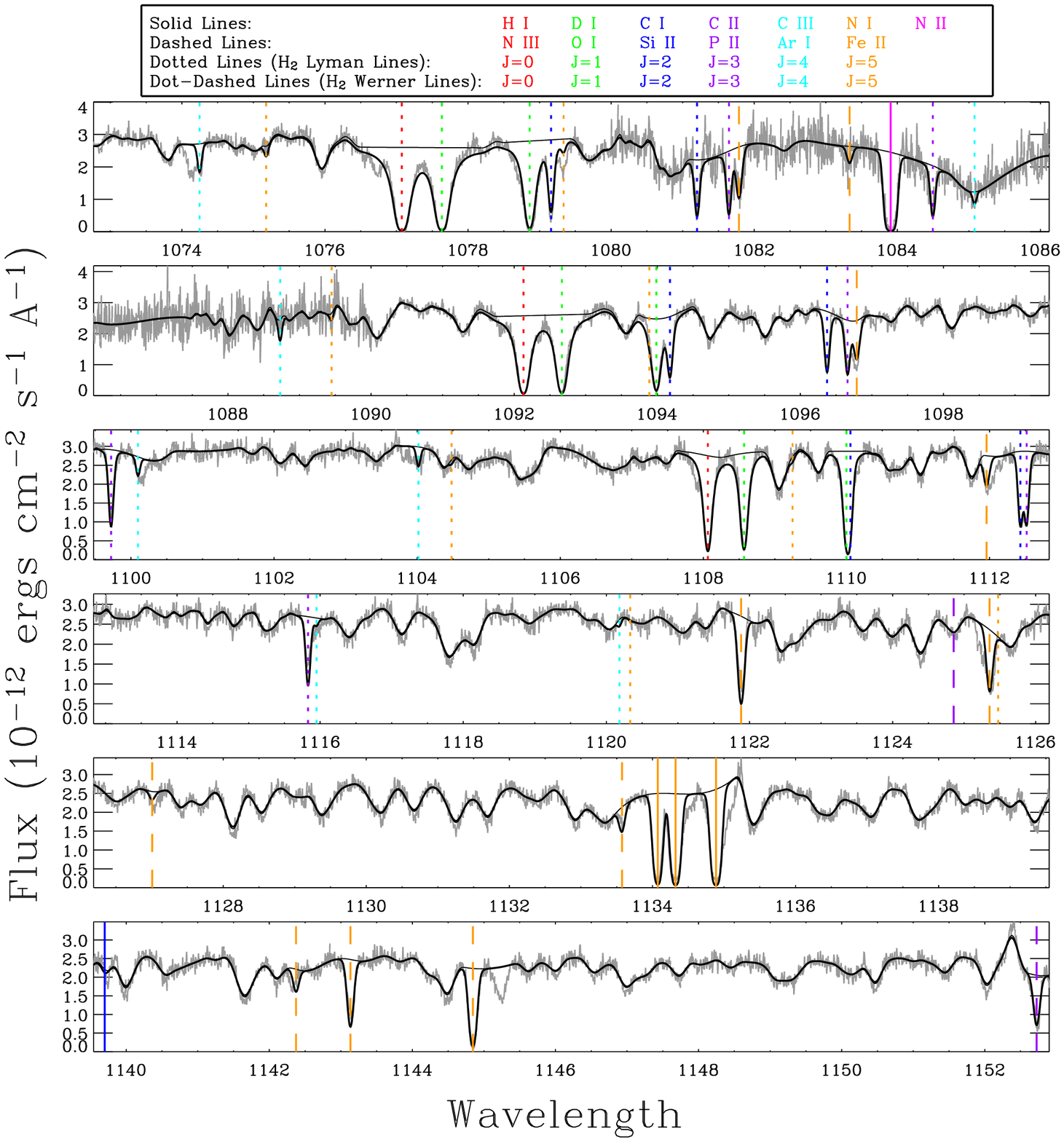}{7.5in}{0}{90}{90}{-390}{0}
\caption{(continued)}
\end{figure}

\clearpage

\begin{figure}
\plotfiddle{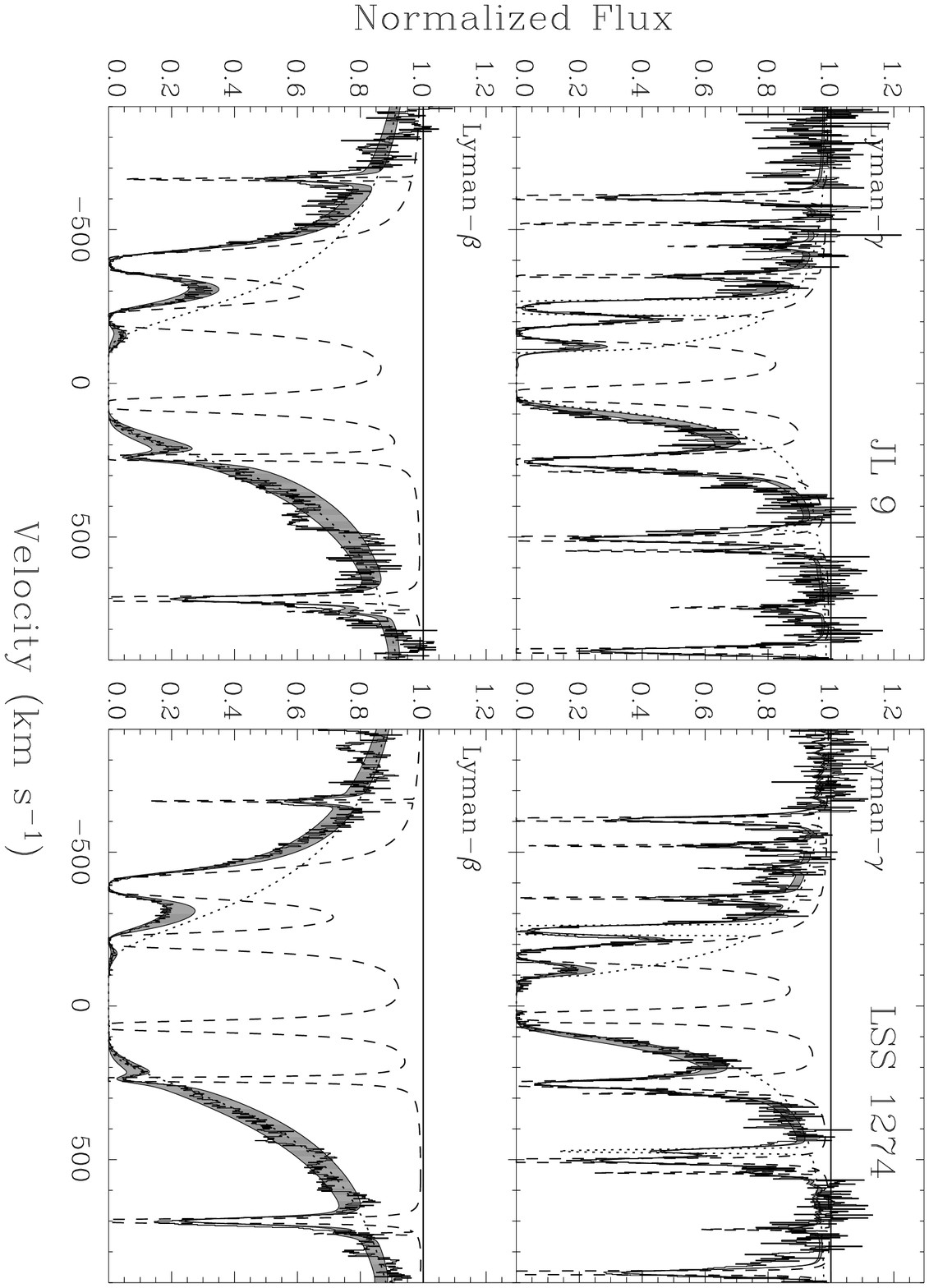}{3.5in}{90}{75}{75}{295}{-10}
\caption{The H~I Ly$\beta$ and Ly$\gamma$ lines of JL~9 and
  LSS~1274, which are the only H~I lines that have substantial
  damping wings that can be used to measure an accurate H~I column density.
  Fits to the data are shown, which are actually part of global fits like
  that in Fig.~1.  The dotted line in each panel is the absorption from
  all the atomic lines (including H~I and D~I) and the dashed line is the
  H$_{2}$ absorption, both lines shown prior to convolution with the
  instrumental LSF.  The shaded region shows the range of profile fits
  after convolution, defined by the $\pm 2\sigma$ range of acceptable H~I
  column densities listed in Table~3.}
\end{figure}

\clearpage

\begin{figure}
\plotfiddle{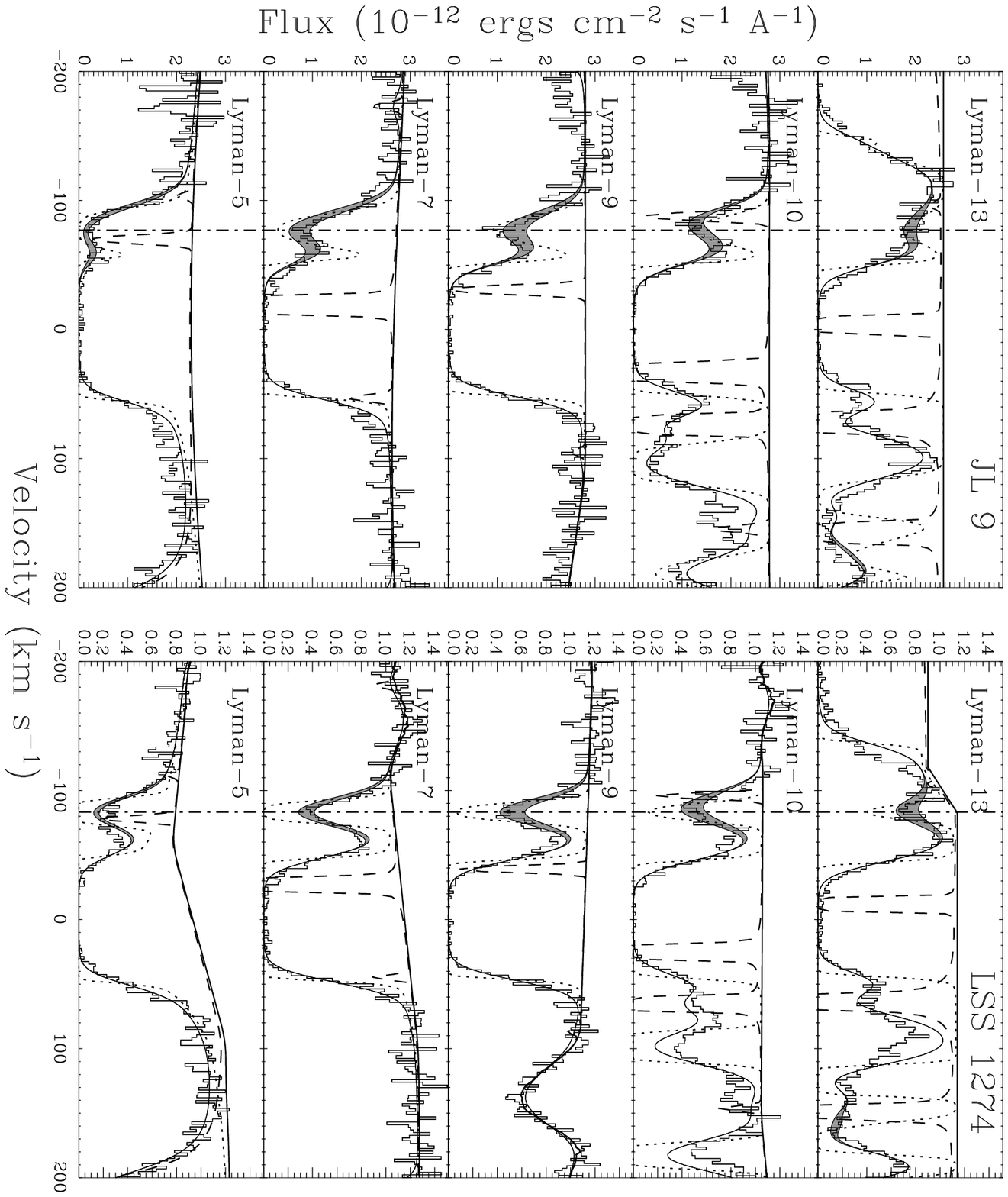}{4.5in}{90}{85}{85}{330}{-10}
\caption{A closeup of all useful D~I lines (marked by dot-dashed line) in
  the FUSE spectra of JL~9 and LSS~1274, plotted on a velocity scale
  centered on the H~I lines bordering D~I.  Fits to the data are shown,
  which are actually part of global fits to the full spectra, as in Fig.~1.
  The dotted line in each panel is the absorption from all the atomic lines
  (including H~I and D~I) and the dashed line is the H$_{2}$ absorption,
  both lines shown prior to convolution with the instrumental profile.  The
  shaded region shows the range of profile fits after convolution, defined
  by the $\pm 2\sigma$ range of acceptable D~I column densities listed in
  Table~3.}
\end{figure}

\begin{figure}
\plotfiddle{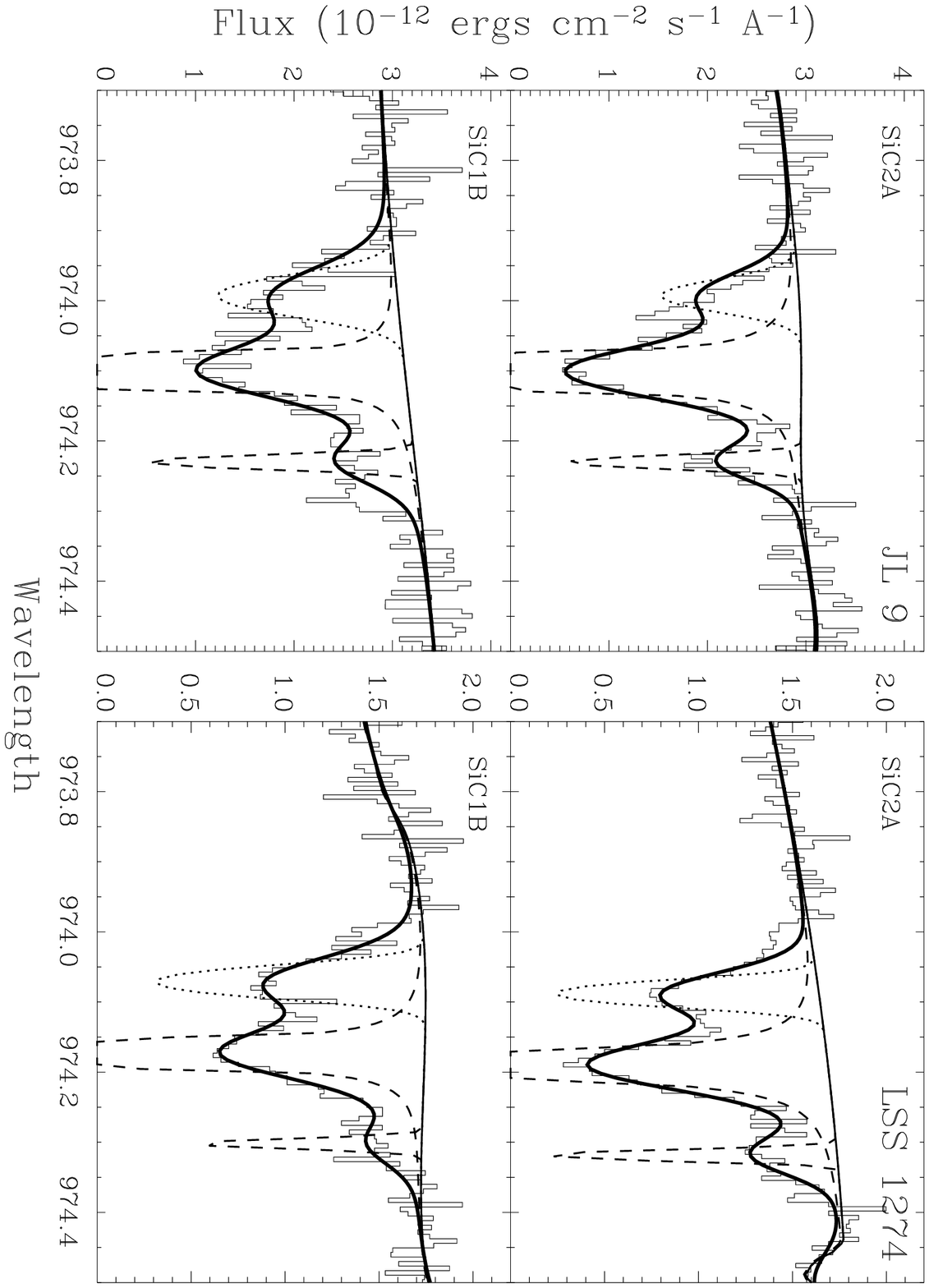}{3.5in}{90}{75}{75}{295}{-10}
\caption{Fits to the blended O~I 974.07~\AA, H$_{2}$(J=2) 974.16~\AA,
  and H$_{2}$(J=5) 974.28~\AA\ lines.  The dotted (O~I) and dashed (H$_{2}$)
  lines show the individual absorption components, and the thick solid lines
  show the total absorption after convolution with the instrumental profile.
  The fits are shown for both the SiC2A and SiC1B segments.}
\end{figure}

\begin{figure}
\plotfiddle{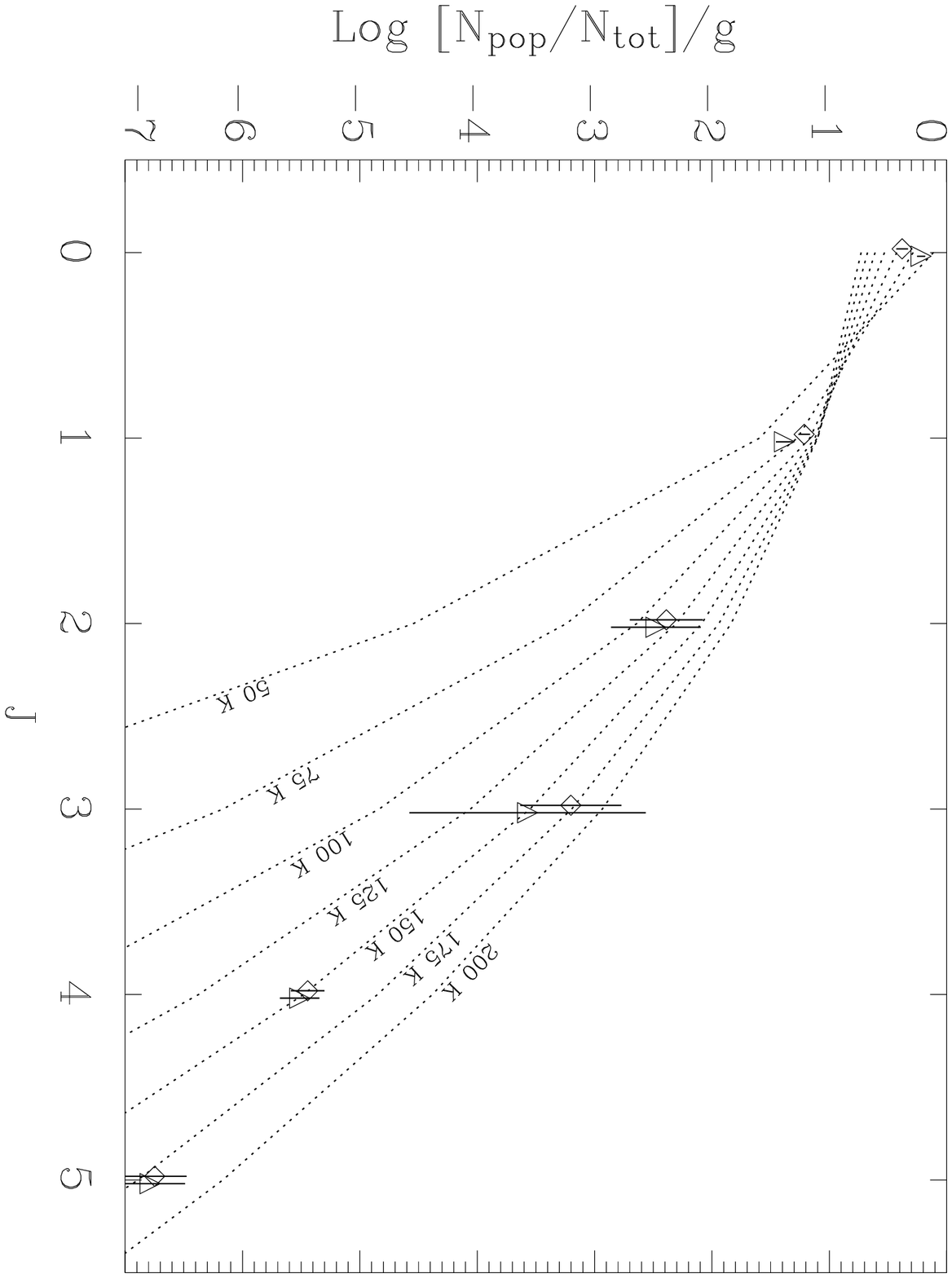}{3.5in}{90}{75}{75}{290}{0}
\caption{The relative H$_{2}$ level populations (normalized by the
  statistical weight $g$) for the JL~9 (diamonds) and LSS~1274 (triangles)
  lines of sight.  Also shown are curves indicating thermal populations
  for temperatures of $50-200$~K.}
\end{figure}

\begin{figure}
\plotfiddle{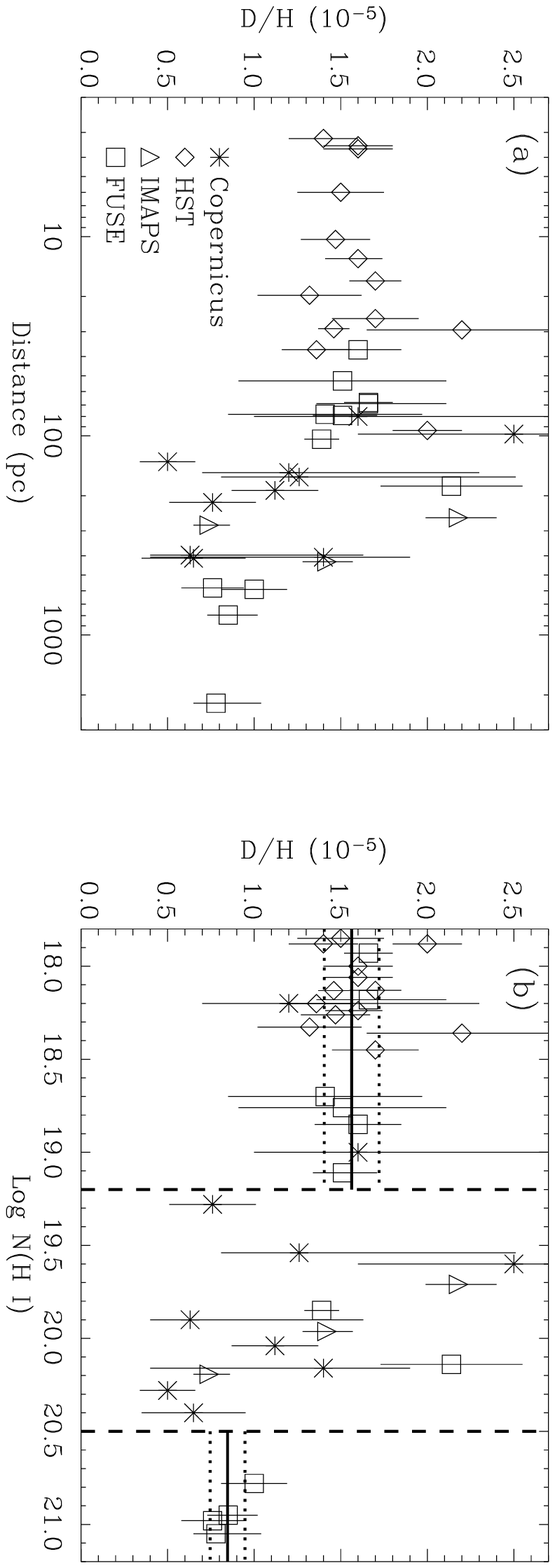}{3.0in}{90}{75}{75}{290}{-180}
\caption{(a) D/H plotted versus line-of-sight distance, using the D/H
  measurements listed in Table~4.  Different symbols are used for different
  sources of the D~I measurement.  (b) D/H plotted versus line-of-sight
  H~I column density.  The symbols are the same as in (a).  D/H appears to
  be constant for $\log N({\rm H~I})<19.2$ and for $\log N({\rm H~I})>20.5$,
  but with different values of ${\rm D/H}=(1.56\pm 0.16)\times 10^{-5}$ and
  ${\rm D/H}=(0.85\pm 0.10)\times 10^{-5}$, respectively.  These weighted
  means and 1$\sigma$ standard deviations are shown in the
  figure.  For intermediate values of $19.2<\log N({\rm H~I})<20.5$ D/H
  appears to be variable.}
\end{figure}

\end{document}